\documentclass[a4paper,11pt,twoside]{article}

\usepackage[latin1]{inputenc} 
\usepackage[T1]{fontenc} 
\newlength{\minipagewidth}
\usepackage{a4wide}
\usepackage{amsmath,amsthm}
\usepackage{amssymb}
\usepackage{a4wide}
\usepackage{graphicx}
\usepackage{color}
\usepackage{epic}
\usepackage{color}

\newcommand{\W}{\mathcal W}
\newcommand{\ph}{\varphi}
\newcommand{\E}{\mathbb E}

\newcommand{\pare}[1]{ \left(#1\right) }
\newcommand{\Frac}[2] {\displaystyle{\frac{#1}{#2}}}

\newenvironment{system}
{ \left \{
  \begin{array}{lcl} }
{  \end{array}
  \right.}

\def\R{\mathbb{R}}
\def\I{\mathrm{Id}}

\newtheorem{The}{Theorem}[section]

\newtheorem{Pro}[The]{Proposition}

\newtheorem{Lem}[The]{Lemma}  

\newtheorem{remark}[The]{\bf Remark}

\title{Computation of free energy differences through nonequilibrium stochastic dynamics:\\ the reaction coordinate case}
\author{Tony Leli\`evre$^{1}$, Mathias Rousset$^{1}$, Gabriel Stoltz$^{1,2}$\\
\footnotesize{1: CERMICS, Ecole Nationale des
  Ponts et Chauss\'ees (ParisTech),}\\ \footnotesize{6 \& 8 Av. Pascal,
  77455 Champs-sur-Marne, France.} \\
\footnotesize{2: CEA/DAM Ile-de-France,
BP 12, 91680 Bruy\`eres-le-Ch\^atel, France.}\\
\footnotesize{ \{lelievre,stoltz\}@cermics.enpc.fr 
  \hspace{1cm} mathias.rousset@polytechnique.org} 
}

\begin{document}

\maketitle

\begin{abstract}
The computation of free energy differences through an exponential
weighting of out-of-equilibrium paths (known as the Jarzynski equality~\cite{jarz,jarzPRL})
is often used for transitions between states described by an external
parameter $\lambda$ in the Hamiltonian. We present here an extension to
transitions between states defined by different values of some reaction
coordinate, using a projected Brownian dynamics. In contrast with other
approaches (see {\it e.g.}~\cite{PKTS03}), we use a projection rather than a constraining potential to let the
constraints associated with the reaction coordinate evolve. We show how to use the Lagrange multipliers associated with these constraints to compute the
work associated with a given trajectory.
Appropriate discretizations are proposed. Some numerical results demonstrate the applicability of the method for the computation of free energy difference profiles.
\end{abstract}


\medskip


The free energy of a system is a quantity of paramount importance in
 statistical physics. It is of the form
\begin{equation}
\label{free_energy_intro}
F = - \beta^{-1} \ln Z,
\end{equation}
where $\beta = 1/ (k_{\rm B}T)$ ($T$ denotes the temperature and $k_{\rm{B}}$ the
Boltzmann constant) and $Z$ is the partition function
\begin{equation}
\label{eq:Z}
Z = \int_{\Sigma} \exp (-\beta V) \, d \mu
\end{equation}
of the Boltzmann (or Gibbs) measure $\exp (-\beta V)  d \mu $. In this expression, the function
$V \equiv V(q)$ is the potential energy of the system (denoting by $q$
the position vector) and $\mu$ is a reference positive measure with
support $\Sigma$. The space $\Sigma$ is the configuration space of the
system. We will consider here that $\Sigma$ is a submanifold of
$\mathbb{R}^{3N}$, but all the results extend to the case when $\Sigma$ is a submanifold of~$\mathbb{T}^{3N}$ (the $3N$-dimensional torus, which arises when using periodic boundary
conditions). The statistics of the system are completely
defined by $(V,\mu)$.

In most cases, $(V,\mu)$ is labeled using a
$d$-dimensional parameter~$z$ (with $d \ll 3N$) which characterizes the system at some coarser level. The parameter $z$ can be independent of the current configuration of the system. In this case, only the expression of the potential~$V$ depends on
the parameter, so that the associated switching has sometimes been called 'alchemical
transition'. Some examples of such parameters are the intensity of
an external magnetic field for a spin system, or the temperature for a simulated annealing process. However, it is often the case that the parameter~$z$ labels submanifolds
of the configuration space, through level sets~$\Sigma_z=\{ \, \xi(q)=z
\, \}$ of some function $\xi$. The function $\xi$ is called a `reaction
coordinate'. In this case, $\mu$ (especially the support of $\mu$) depends on $z$ and is
defined using the orthogonal projection from~$\mathbb{R}^{3N}$
or~$\mathbb{T}^{3N}$ to~$\Sigma_z$ (this will be made precise in Section~\ref{section_TI_def}). Standard examples of reaction coordinates are bond lengths or dihedral angles in a molecule.

The absolute free energy~(\ref{free_energy_intro}) can be computed only for certain systems, such as ideal gases, or solids at low temperature (resorting to the phonon spectrum)~\cite{RS02}. However, in many applications, the quantity of interest is the free
energy \emph{difference} between an initial and a final state (characterized by
two different values of the parameter $z$). The free energy difference
profiles indeed give
information about the relative stabilities of several species, as well as their transition kinetics.
The free energy differences are much more
amenable to compute than the absolute free energy. Classical techniques to this end fall within three
main classes. The first one, dating back to Kirkwood~\cite{Kirk35}, is
{\it thermodynamic integration}, which mimics the
quasi-static evolution of a system as a succession of equilibrium
samplings, which amounts to an infinitely slow switching between the
initial and final states. The second one, the {\it free energy perturbation
method}, was introduced by Zwanzig~\cite{Zwanz54}. It recasts free energy
differences as a phase-space integral, so that usual sampling techniques
can be employed. Notice also that there exist many refinements for those
two classes of techniques, such as umbrella sampling~\cite{TV77}. The last and most recent
class of methods uses dynamics arising from a switching at a finite rate. This can be done 
using {\it nonequilibrium dynamics} (the so-called fast growth methods) with a suitable exponential reweighting, 
as introduced by Jarzynski in~\cite{jarz}. 
Notice that the thermodynamic integration and free energy perturbation methods can 
be seen respectively as the limits of infinitely slow and fast switching of  
nonequilibrium dynamics, at least formally. Instead of being imposed {\it a priori}, this
switching may also arise as the result of an equilibrium sampling, using
for example the Adaptive Biasing Force
technique~\cite{DP01,HC04} or metadynamics~\cite{ILP03}. In those cases, the system 
is progressively forced to leave regions where the sampling of the reaction coordinate 
has been completed. 

%
It is still a matter of debate which method is the most efficient. 
While some results show that fast growth methods can be competitive in some 
situations~\cite{HJ01}, other studies disagree~\cite{ODG05}. The results
of~\cite{ODG05} indeed indicate that even with the use of efficient
path sampling techniques (see also~\cite{Stuart,Sun,YZ04}), fast growth
methods do not outperform conventional methods such as umbrella sampling
or thermodynamic integration (at least in a number of typical
cases). However, general conclusions about the efficiency of fast
growth methods are still to be drawn, depending on the cases under
consideration. We believe that there is room for improvements of this
relatively new method ({\it e.g.} by optimizing the switching
schedule~\cite{RD03}). Let us also mention that this method is
straightforward to parallelize and naturally provides with {\it a
  posteriori} error bounds {\it via} the central limit theorem, since it
involves many independent trajectories.

Most methods to compute free energy differences are well suited to the alchemical 
transition setting, but do not straightforwardly extend to the reaction
coordinate setting. This latter case is the focus throughout this article.
 In this case, the methods described above require to consider dynamics
 restricted to the submanifold $\Sigma_z$. For computations using
Hamiltonian dynamics, we refer for example to~\cite{CCHK89,RD03}. In the stochastic case, thermodynamic integration in the reaction coordinate case using projected stochastic 
dynamics has recently been put on a firm grounding~\cite{LLV05,EV04}. On
the other hand, stochastic nonequilibrium dynamics {\it \`a la}
Jarzynski in the reaction coordinate case was, to our knowledge, not
studied mathematically.  It is the aim of this paper to perform such a
study and to present a methodology to compute free energy differences in
this framework.

Nonequilibrium computations of free energy differences in the
reaction coordinate setting using stochastic dynamics have until now
used soft constraints to switch between the initial state centered on
the submanifold $\{ \xi(q) = z_0 \}$ and the final state centered on $\{
\xi(q) = z_1 \}$. Steered molecular dynamics techniques use for example
a penalty term $K (\xi(q)-z)^2$ in the energy of the
system~\cite{PKTS03} (with $K$ large) to 'softly' constraint the system
to remain close to the submanifold $\{ \xi(q)-z=0 \}$, and varying the
value $z$ from 0 to 1 in a finite time~$T$.  It is shown in~\cite{HS01}
how to use such a biasing potential to exactly compute free energy differences
(even for a finite $K$), which is of particular interest for
experimental studies. From a computational viewpoint
however, it is expected that large values of $K$ require
small integration time steps.
Moreover, it is observed in practice that the statistical fluctuations increase with larger~$K$ (see~\cite{PKTS03}). Instead, we propose to replace the stiff constraining potential $K (\xi(q)-z)^2$ by a projection onto the submanifold $\{ \xi(q)-z=0 \}$. This situation is reminiscent of the case of molecular constraints, that can be enforced using a stiff penalty term, or more elegantly and often more efficiently, using some 
projection of the dynamics involving Lagrange multipliers. This is the
spirit of the well known SHAKE algorithm~\cite{RCB77}.

We propose a nonequilibrium stochastic dynamics and an equality that allow to
compute free energy differences between states defined by different
values of a reaction coordinate. The dynamics relies on a projection
onto the current submanifold at each time step, and we use the Lagrange multipliers associated with this projection to estimate the free energy difference. More precisely, we use the difference between these Lagrange multipliers and the external forcing term required for the finite time switching (see for example the discretization~(\ref{mean_force_evolution})). The main results of the paper are the Feynman-Kac equality of
Theorem~\ref{th:FK} (which extends the proof of~\cite{HS01} to hard constraints), as well as the associated discretizations~(\ref{eq:num_work}) and~(\ref{eq:jarz_estimator}).

The method we propose forces the system to pass free energy
barriers, and thus enables free energy difference computations for
metastable systems. Of course the reliability of the algorithm crucially
depends on the choice of the reaction coordinate, which represents the
essential degrees of freedom. The reaction
coordinate should be rich enough in order to adequately describe the
configuration paths of the system from the initial state to the final
state. The determination of the essential degrees of freedom of a system
is a very important problem, which is not the focus of this work. Thus,
in the following, we suppose that a ``good'' reaction coordinate is
given, and we are interested in the computation of free energy
differences associated with this reaction coordinate.

Let us also notice that some recent refinements of nonequilibrium dynamics to compute 
free energy differences, especially path sampling techniques~\cite{YZ04} and Interacting 
Particle Systems approaches~\cite{RS06} (which equilibrate the nonequilibrium dynamics 
through some birth/death process based on the current work), can be extended to the 
reaction coordinate setting using the techniques we present
here. Moreover, we restrict 
ourselves to the so-called overdamped Langevin dynamics, but it is
possible to extend these results to the usual Langevin 
dynamics (this is a work in progress).

The paper is organized as follows. In Section~\ref{section_TI}, the
thermodynamic integration setting is outlined in the reaction coordinate
case. Section~\ref{section_jarz} then extends the method to
nonequilibrium dynamics. Adapted numerical schemes are
proposed in Section~\ref{section_discretization}, and some numerical
results assessing the correctness of the method are presented in
Section~\ref{section_res_num}.  For clarity, we present the
method in the case of a one-dimensional reaction coordinate and postpone
until Appendix~\ref{sec:multi} the proofs and the expressions for the multi-dimensional case.




\section{Equilibrium computation of free energy differences}
\label{section_TI}

The aim of this section is to introduce the definitions of the free
energy and the mean force, and to recall how thermodynamic
integration is used to compute free energy differences. The computation of the
mean force is based on projected stochastic differential equations
(SDE). These SDEs will also be used for the discretization of Jarzynski equality in Section~\ref{section_jarz}. This section mainly reviews results of~\cite{LLV05}.

\subsection{Free energy and mean force}
\label{section_TI_def}

In the following, we denote by $\mathcal{M}\subset\R^{3N} $ the
configuration space of the system when no parameter $z$ is involved. The
state of the system is characterized by the value of a reaction
coordinate $\xi \, : \, \mathcal{M} \to [0,1]$. The function $\xi$ is
supposed to be smooth and such that $\nabla \xi(q)
\not = 0$ for all $q \in \mathcal{M}$. For a given value $z \in [0,1]$,
we denote by $\Sigma_z$ the submanifold
\begin{equation}\label{eq:sigma_z}
\Sigma_z=\{ \ q \in {\cal M}, \, \xi(q)=z \ \}
\end{equation}
and we assume that $\bigcup_{z \in [0,1]} \Sigma_z \subset \mathcal{M}$. 
For each point $q \in \Sigma_z$, we also introduce the orthogonal
projection operator $P(q)$ onto the
tangent space to $\Sigma_z$ at point $q$ defined by:
\begin{equation}
\label{eq:P}
P(q)=\I-\frac{\nabla \xi \otimes \nabla \xi}{|\nabla \xi|^2}(q),
\end{equation}
where $\otimes$ denotes the tensor product. The orthogonal projection
operator on the normal space to $\Sigma_z$ at point $q$ is defined by $P^\perp(q)=\I-P(q)$.

The free energy is then defined as
\begin{equation}
\label{eq:F_z}
F(z)=-\beta^{-1} \ln \left ( Z_z \right ),
\end{equation}
with
\begin{equation}
\label{eq:Z_z}
Z_z=\int_{\Sigma_z} \exp( -\beta V ) \, d\sigma_{\Sigma_z},
\end{equation}
where for any submanifold $\Sigma$ of $\R^{3N}$, $\sigma_{\Sigma}$
denotes the Lebesgue measure induced on $\Sigma$ as a submanifold of
$\R^{3N}$. The associated Boltzmann probability measure is
\begin{equation}
\label{eq:mu_z}
d\mu_{\Sigma_z} = Z_z^{-1} \exp( -\beta V ) \, d\sigma_{\Sigma_z}.
\end{equation}

\begin{remark}[On the definition of the free energy]\label{rem:def_F}
Two comments are in order about formula~(\ref{eq:F_z}). First, this
formula is valid up to an additive constant, which is not important when
considering free energy differences. Second,
the potential $V$ in~(\ref{eq:Z_z}) may be a
potential different from the actual potential seen by the particles. More
precisely, if the particles evolve in a potential $V$, the standard
definition of the free energy in the physics and chemistry literature is~(\ref{eq:F_z})
with
\[
Z_z=\int \exp( -\beta V ) \, \delta_{\xi(q)-z},
\]
where $\delta_{\xi(q)-z}$ is a measure supported by $\Sigma_z$ and
defined by: for all test functions $\phi$,
$$\int \phi(q)\delta_{\xi(q)-z} = \int_{\Sigma_z} \phi |\nabla \xi|^{-1} \, d\sigma_{\Sigma_z}.$$
This amounts to considering~(\ref{eq:F_z})--(\ref{eq:Z_z}) with $V$
replaced by an effective potential
$V+\beta^{-1} \ln |\nabla \xi |$ (see Remark~\ref{rem:def_F_multi} for
the case of a multi-dimensional constraint). Since the results we
present in this paper hold irrespective of the physical signification of
the potential $V$, we may assume without loss of mathematical generality
that the free energy is indeed given by~(\ref{eq:F_z})--(\ref{eq:Z_z}).
Let us emphasize that, in practice, the cumbersome computation of the
gradient of the additional term $\beta^{-1} \ln |\nabla \xi |$ in the
modified potential (which intervenes in the projected SDEs we use,
see~(\ref{SDE_discretisee_implicite})--(\ref{SDE_discretisee_explicite})
or~(\ref{SDE_evolution_implicite})--(\ref{SDE_evolution_explicite})) can
be avoided resorting to some finite differences, as explained in~\cite{LLV05}.
\end{remark}

Using the co-area formula (see~(\ref{eq:co_area}) and Proposition~\ref{prop:F_multi} for a proof
in the multi-dimensional case), it is possible to derive the following
expression of the derivative of the free energy $F$ with respect to $z$
(the so-called \emph{mean force}) (see~\cite{OB98,SC98}):
\begin{equation}
\label{eq:mean_force}
F'(z)=Z_z^{-1} \displaystyle{\int_{\Sigma_z}  \frac{\nabla \xi}{|\nabla \xi|^2} \cdot (\nabla V +
  \beta^{-1} H) \exp( -\beta V ) d\sigma_{\Sigma_z}},
\end{equation}
where
\begin{equation}
H=-\nabla \cdot \left( \frac{\nabla \xi}{|\nabla \xi|} \right) \frac{\nabla \xi}{|\nabla \xi|}
\end{equation}
is \emph{the mean curvature vector field} of the surface $\Sigma_z$.
The free energy can thus be expressed as an average with respect to~$\mu_{\Sigma_z}$:
\begin{equation}
\label{eq:mean_force_mu}
F'(z)= \displaystyle{\int_{\Sigma_z}  f(q) d\mu_{\Sigma_z}}(q),
\end{equation}
where $f$ is the local mean force defined by:
\begin{equation}
\label{eq:loc_force}
f=\frac{\nabla \xi}{|\nabla \xi|^2} \cdot (\nabla V + \beta^{-1} H).
\end{equation}
In next section, we will explain how it is possible
to compute this average with respect to~$\mu_{\Sigma_z}$, without
explicitly computing $f$, by using projected SDEs.
This avoids in particular the computation of the mean curvature vector $H$
which involves second-order derivatives of~$\xi$.


The principle of \emph{thermodynamic integration} is to recast the free energy difference 
\begin{equation}
\Delta F(z)=F(z)-F(0)
\end{equation}
between two reaction coordinates $0$ and $z$ as an integral over the
mean force:
\begin{equation}
\label{TI}
\Delta F(z) = \int_{0}^{z} F'(y) \ dy.
\end{equation}

Therefore, in practice, thermodynamic integration computation of
free-energy is as
follows. First, the free energy difference $\Delta F(z)$ is estimated
using quadrature formulae for the integral in~(\ref{TI}), such as for example a Gauss-Lobatto scheme:
\[
\Delta F(z) \simeq \sum_{i=0}^K  \omega_i F'(y_i)
\]
where the points $\{ y_0, y_1, \dots, y_K \}$ are in $[0,z]$ and $\{
\omega_0, \omega_1, \dots,  \omega_K \}$ are their associated weights. Second, the
derivatives $F'(y_i)$ are computed as canonical averages over the
submanifolds $\Sigma_{y_i}$, using projected SDEs (see next section).

To obtain a free-energy profile (and not only a free-energy difference
for a fixed final state),
it is possible to approximate the function $\Delta F(z)$ on the interval
$[0,1]$ by a polynomial. This can be done for example by interpolating
the derivative $F'$ by splines, and integrating the resulting function
(consistently with the normalization $\Delta F(0) = 0$).

\subsection{Projected stochastic differential equations}

In this section, we explain how to compute the mean force $F'(z)$
defined by~(\ref{eq:mean_force}) using projected SDEs, for a fixed
parameter $z$. We consider the solution $Q_t$ to the following SDE:
\begin{equation}\label{eq:Q}
\begin{cases}
Q_0 \in \Sigma_z, \\
dQ_t= - P(Q_t) \nabla V(Q_t) \, dt + \sqrt{2 \beta^{-1}} P(Q_t) \circ dB_t,
\end{cases}
\end{equation}
where $B_t$ is the standard $3N$-dimensional Brownian motion and $\circ$ denotes the Stratonovich product.
It is possible (see~\cite{LLV05}) to check that $\mu_{\Sigma_z}$ is an invariant probability measure associated with the
SDE~(\ref{eq:Q}). Under suitable assumptions, which we assume in the rest of the section, on the potential $V$ and
the surface $\Sigma_z$, the process~$Q_t$ is ergodic with respect
to~$\mu_{\Sigma_z}$. Moreover, the SDE~(\ref{eq:Q}) can be rewritten in the
following way:
\begin{equation}
\label{eq:Q_proj}
dQ_t= - \nabla V(Q_t) \, dt + \sqrt{2 \beta^{-1}} dB_t + \nabla \xi(Q_t)
d \Lambda_t,
\end{equation}
where $\Lambda_t$ is a real valued process, which can be interpreted as
the Lagrange multiplier associated with the constraint
$\xi(Q_t)=z$ (see the discretization in
Section~\ref{num_constrained_SDE}). This process can be decomposed into two parts:
\begin{equation}
d\Lambda_t=d\Lambda^{\rm m}_t + d\Lambda^{\rm f}_t.
\end{equation}
The so-called martingale\footnote{For our purposes, it is enough to think of a
  martingale as an It\^o integral with respect to the Brownian motion
  $(B_t)_{t \geq 0}$.}  part  $\Lambda^{\rm m}_t$ (whose fluctuation is
of order
$\sqrt{\Delta t}$ over a timestep $\Delta t$) is
\begin{equation}
d\Lambda^{\rm m}_t=- \sqrt{2 \beta^{-1}} \frac{\nabla \xi}{ |\nabla \xi|^2}(Q_t) \cdot dB_t,
\end{equation}
where $\cdot$ implicitly denotes the It\^o product. The so-called bounded
variation part $\Lambda^{\rm f}_t$ (whose fluctuation is
of order $\Delta t$ over a timestep $\Delta t$) is
\begin{equation}
d\Lambda^{\rm f}_t= \frac{\nabla \xi}{ |\nabla \xi|^2}(Q_t) \cdot \nabla V(Q_t) \, dt +
\beta^{-1} \frac{\nabla \xi}{ |\nabla \xi|^2}(Q_t) \cdot H(Q_t) \, dt= f(Q_t) \, dt,
\end{equation}
$f$ being the local mean force defined above
by~(\ref{eq:loc_force}). Thus, since $Q_t$ is ergodic with respect to
$\mu_{\Sigma_z}$ the mean force can be obtained as a mean
over the Lagrange multiplier $\Lambda_t$:
\begin{Pro}\label{prop:F} The mean force is given by:
\begin{equation}
\label{eq:F'}
F'(z) =  \lim_{T \to \infty} \frac{1}{T} \int_0^T d\Lambda_t=  \lim_{T
  \to \infty} \frac{1}{T} \int_0^T d\Lambda^{\rm f}_t.
\end{equation}
\end{Pro}
Notice that the martingale part $d\Lambda^{\rm m}_t$, which has the
largest fluctuations, has zero mean. In order to reduce the variance, it
is thus numerically convenient to perform the mean over the bounded variation part
$d\Lambda^{\rm f}_t$ rather than over the whole Lagrange multiplier
$d\Lambda_t$ (see Section~\ref{section_discretization}).

We refer to~\cite{LLV05} for a proof of Proposition~\ref{prop:F}, as well as for
formulae involving higher dimensional reaction coordinates. Such ideas
have been used for a long time in the framework of Hamiltonian dynamics (see~\cite{OB98,SC98}).

The interest of Equation~(\ref{eq:F'}) is that the SDE~(\ref{eq:Q_proj}) can be
very naturally discretized as explained in
Section~\ref{num_constrained_SDE} below. Then, the
average over a discretized trajectory of the process~$\Lambda_t$
converges to $F'(z)$. This is particularly convenient for numerical
purposes since it does not ask for explicitly computing the local force $f$. 
For further details, we refer to~\cite{LLV05} and
to Section~\ref{num_constrained_SDE}. In next section, we use these
ideas for the computation of the free energy difference given through the Jarzynski equality.



\section{Nonequilibrium stochastic methods in the reaction coordinate case}
\label{section_jarz}

As opposed to quasistatic methods where the free energy difference between an initial state and 
a final state is
expressed by~(\ref{TI}), in nonequilibrium methods, the free energy difference is expressed using a Feynman-Kac average over nonequilibrium 
paths~\cite{jarz,HS01,RS06}
\begin{equation}\label{e:jarz}
\Delta F(1)=F(1)-F(0)=- \beta^{-1} \ln\mathbb{E}\left({\rm e}^{-\beta \W(T)} \right),
\end{equation}
where $\W(T)$ denotes the total work exerted along a
nonequilibrium path $(Q_{t},z(t))_{t\in[0,T]}$, with $z(0)=0$ and
$z(T)=1$.

We wish here to extend the Feynman-Kac formula derived in~\cite{HS01} for
a parameter~$z$ which appears only in the potential $V$, to the
reaction coordinate case, where $z$ labels submanifolds~$\Sigma_z$
(defined by Equation~(\ref{eq:sigma_z})) of the state space. To this end,
we need to make precise the evolution of the constraints.

We consider a ${\mathcal C}^1$ path $z \, : \, [0,T] \to [0,1]$ of values of
the reaction coordinate $\xi$, with $z(0)=0$, and $z(T)=1$. Recall that
the associated family of submanifolds of admissible configurations is
denoted by $$\Sigma_{z(t)}=\left\{q\in \mathcal{M}, \, \xi(q)=z(t)
\right\},$$ and that the associated Boltzmann probability measures are $$d\mu_{\Sigma_{z(t)}} = Z_{z(t)}^{-1}\exp(-\beta V ) d\sigma_{\Sigma_{z(t)}}.$$
We construct a diffusion $(Q_{t})_{t \in
  [0,T]}$ so that $Q_t \in \Sigma_{z(t)}$ for all $t \in [0,T]$ and $(Q_{t})_{t \in
  [0,T]}$ satisfies the following properties (see Section~\ref{sec:diff}
for a more rigorous formulation):
\begin{itemize}
\item $Q_{0} \sim \mu_{\Sigma_{z(0)}}$,
\item For all $t\in [0,T]$, $Q_{t+dt}$ is the orthogonal projection on
  $\Sigma_{z(t+dt)}$ of the position obtained by the unconstrained displacement: $Q_{t}-\nabla V (Q_{t}) dt + \sqrt{2\beta^{-1}}d B_{t} $.
\end{itemize}
To each realization of this process, a work $\W(t)$ can be associated as
$$\W(t)=\int_{0}^{t}f(Q_{s}) z'(s)ds,$$
where $f$ is the local mean force defined above
by~(\ref{eq:loc_force}). Then, we prove that the Feynman-Kac
formula~(\ref{e:jarz}) holds for the free energy $F$ associated with the
reaction coordinate and defined by~(\ref{eq:F_z}).  Notice that, at
least formally, in the
limit of an infinitely slow switching from $z(0)=0$ to $z(T)=1$, Formula~(\ref{e:jarz}) corresponds to
the thermodynamic integration formula~(\ref{TI}). Formula~(\ref{e:jarz})
enables the computation of free energy difference at arbitrary rates,
through a correction consisting in a reweighting of the nonequilibrium paths.

The rest of this section is organized as follows. In Section~\ref{sec:diff}, we make precise
the process $Q_t$ we consider. Then, in Section~\ref{sec:FK}, we state
the Feynman-Kac formula~(\ref{e:jarz}) for a one-dimensional reaction
coordinate. We recall that the formulae for the general case involving
higher dimensional reaction coordinates, as well as the main proofs, are presented in Appendix~\ref{sec:multi}.


\subsection{The nonequilibrium projected stochastic dynamics}
\label{sec:diff}

The considered diffusion reads, in the Stratonovich setting:
\begin{equation}
\label{e:diff}
\begin{system}
Q_{0}  &\sim & \mu_{\Sigma_{z(0)}},\\
d Q_{t} & = & -P(Q_{t})\nabla V (Q_{t}) dt+\sqrt{2\beta^{-1}} P(Q_{t}) \circ d B_{t}+\nabla \xi(Q_{t}) \, d \Lambda^{\rm ext}_{t}, \\
d \Lambda^{\rm ext}_t & = & \displaystyle{\frac{z'(t)}{|\nabla \xi(Q_{t})|^2}  dt}.
\end{system}
\end{equation}
With a view to the discretization of $Q_t$, let us notice that $Q_{t}$ can be characterized by the following property:
\begin{Pro}\label{prop:Qt}
The process $Q_t$ solution to~\eqref{e:diff} is the only
  It\^o process satisfying for some real-valued adapted It\^o process
  $(\Lambda_{t})_{t\in[0,T]}$:
\begin{equation*}
\begin{system}
Q_{0}  &\sim & \mu_{\Sigma_{z(0)}},\\
d Q_{t} & = & -\nabla V (Q_{t}) dt+\sqrt{2\beta^{-1}} d B_{t}+\nabla \xi(Q_{t}) \, 
d \Lambda_{t}, \\
\xi(Q_{t}) & = & z(t).
\end{system}
\end{equation*}
Moreover, the process $(\Lambda_{t})_{t\in[0,T]}$ can be decomposed as
\begin{equation}\label{e:lambdadec}
\Lambda_{t}=\Lambda^{\rm m}_{t}+\Lambda^{\rm f}_{t} + \Lambda^{\rm ext}_{t},
\end{equation}
with the martingale part
\[
d \Lambda^{\rm m}_{t}= -\sqrt{2\beta^{-1}} \frac{\nabla \xi}{|\nabla \xi|^2}(Q_{t}) \cdot dB_{t},
\]
the local force part (see~(\ref{eq:loc_force}) for the definition of $f$)
\begin{equation}\label{eq:Lambda^f}
d \Lambda^{\rm f}_t =\frac{\nabla \xi}{|\nabla \xi|^2}(Q_{t}) \cdot \left( \nabla V (Q_{t}) \, dt + \beta^{-1} H(Q_{t}) \right) dt = f(Q_t) \, dt, \\
\end{equation}
and the external forcing (or switching) term
\[
d \Lambda^{\rm ext}_t = \frac{z'(t)}{|\nabla \xi(Q_{t})|^2} \, dt.
\]
\end{Pro}
The proof of Proposition~\ref{prop:Qt} is easy and consists in computing $d\xi(Q_t)$ by It\^o's calculus and
identifying the bounded variation and the martingale parts of the
stochastic processes.

The difference with the projected stochastic differential
equation~(\ref{eq:Q}) considered in the thermodynamic integration setting is
that the out-of-equilibrium evolution of the constraints~$z(t)$ creates a drift $\nabla \xi(Q_{t}) \, d \Lambda^{\rm ext}_{t}$ along the reaction coordinate. This drift can be interpreted as an external forcing required for the switching
to take place at a finite rate, and must be subtracted from the Lagrange multiplier $\Lambda_t$ in order to obtain a correct expression for the work $\W(t)$ involved in the Feynman-Kac fluctuation
equality (see Equations~(\ref{mean_force_evolution}) and~(\ref{eq:num_work}) below). This correction is quantitatively important when
the switching is not slow.

\subsection{The Feynman-Kac fluctuation equality}
\label{sec:FK}

Let us define the nonequilibrium work exerted on the diffusion
\eqref{e:diff} by:
\begin{equation}
\label{e:work}
\W(t) = \int_{0}^{t} f(Q_{s}) \, z'(s)\, ds,
\end{equation}
where $f$ is the local mean force defined above
by~(\ref{eq:loc_force}). In practice, the nonequilibrium work~$\W(t)$ can be computed by using the
 local force part $d\Lambda^{\rm f}_t$ (see~(\ref{eq:Lambda^f})), as in the
 thermodynamic integration method (see~(\ref{eq:F'})). Thus, the formula
 we use to compute $\W(t)$ is rather:
\begin{equation}
\label{eq:work_prac}
\W(t) = \int_{0}^{t}  z'(s)\, d \Lambda^{\rm f}_s,
\end{equation}
since $\Lambda^{\rm f}_t$ can be obtained by a natural numerical scheme (see Section~\ref{section_discretization}), 
avoiding the cumbersome computations of the mean curvature vector $H$ in the expression of $f$
(as already explained in Section~\ref{section_TI_def}). 

We can now state the generalization of the Jarzynski nonequilibrium
equality to the case when the switching is parameterized by a reaction coordinate.

\begin{The}[Feynman-Kac fluctuation equality]\label{th:FK}
For any test function $\ph$ and $\forall t \in [0,T]$, it holds
\begin{equation*}
\frac{Z_{z(t)}}{Z_{z(0)}} \int_{\Sigma_{z(t)}}\!\!\!\!\!\! \ph\, d\mu_{\Sigma_{z(t)}} = \mathbb{E} \left ( \ph(Q_{t}){\rm e}^{-\beta \W(t)} \right ).
\end{equation*}
In particular, we have the work fluctuation identity: $\forall t \in [0,T]$,
\begin{equation}
\label{eq:FK}
\Delta F(z(t)) =F(z(t))-F(z(0))= - \beta^{-1} \ln \left( \mathbb{E} \left ( {\rm
      e}^{-\beta \W(t)} \right ) \right).
\end{equation}
\end{The}
As in the alchemical case~\cite{HS01}, the proof follows from a
Feynman-Kac formula. The proof of this theorem is presented
in the general multi-dimensional case in Appendix~\ref{sec:multi} (see Theorem~\ref{th:FK_multi}).


\section{Discretization of the dynamics}
\label{section_discretization}

The main interest of the above formulae~(\ref{TI})--(\ref{eq:F'}) and~(\ref{eq:work_prac})--(\ref{eq:FK}) is that they admit
natural time discretizations. The principle is to use a
predictor-corrector scheme for the associated dynamics~(\ref{eq:Q})
and~(\ref{e:diff}), and to use the Lagrange multiplier $\Lambda_t$
to compute the local mean force $f$.

Section~\ref{num_constrained_SDE} is mainly a review of the results
of~\cite{LLV05} and presents this idea in the
context of thermodynamic integration. Then, we extend the method to the
case of evolving constraints in Section~\ref{sec:disc_evolv_const}.

\subsection{Discretization of the projected diffusion}
\label{num_constrained_SDE}

For the projected SDE \eqref{eq:Q_proj} onto a
submanifold $\Sigma_{z}=\{ \xi(q) -z = 0 \}$, two discretizations  of the
dynamics, extending the
usual Euler-Maruyama scheme, are proposed in~\cite{LLV05}. These
numerical schemes for constrained Brownian dynamics are in the spirit of
the so-called RATTLE~\cite{rattle} and SHAKE~\cite{RCB77} algorithms
classical used for constrained Hamiltonian dynamics, and also related with
the algorithms proposed in~\cite{VB82,AM84,oettinger-94}.

The first one is:
\begin{equation}
\label{SDE_discretisee_implicite}
\left \{ \begin{array}{l}
Q_{n+1} = Q_n - \nabla V(Q_n) \, \Delta t + \sqrt{2 \Delta t \, \beta^{-1}} \, U_n + \Delta\Lambda_{n+1} \, \nabla \xi(Q_{n+1}), \\
\mbox{where $\Delta\Lambda_{n+1}$ is such that $\xi(Q_{n+1}) = z$,}
\end{array} \right.
\end{equation}
where $\Delta t$ is the time step and $U^n$ is a $3N$-dimensional standard Gaussian random vector. Notice
that~(\ref{SDE_discretisee_implicite}) admits a natural variational
interpretation, since $Q_{n+1}$ can be seen as the closest point
on the submanifold $\Sigma_{z}$ to the predicted position $Q_n - \nabla
V(Q_n) \, \Delta t + \sqrt{2 \Delta t \beta^{-1}} \, U_n$. The real
$\Delta\Lambda_{n+1}$ is then the Lagrange multiplier associated with the
constraint $\xi(Q_{n+1}) = z$.

Another possible discretization of \eqref{eq:Q_proj} is
\begin{equation}
\label{SDE_discretisee_explicite}
\left \{ \begin{array}{l}
Q_{n+1} = Q_n - \nabla V(Q_n) \, \Delta t + \sqrt{2 \Delta t \, \beta^{-1}} \, U_n + \Delta\Lambda_{n+1} \, \nabla \xi(Q_n), \\
\mbox{where $\Delta\Lambda_{n+1}$ is such that $\xi(Q_{n+1}) = z$.}
\end{array} \right.
\end{equation}
Although this scheme is not naturally associated with a variational
principle, it may be more practical since its formulation is more
explicit. Notice also that we use the same notation~$\Delta \Lambda_n$ for the Lagrange multipliers for both~(\ref{SDE_discretisee_implicite}) and~(\ref{SDE_discretisee_explicite}) (and later for~(\ref{SDE_evolution_implicite}) and~(\ref{SDE_evolution_explicite})), since all the formulas we state in terms of $\Delta \Lambda_n$ are verified whatever the constrained dynamics.

To solve Equation~(\ref{SDE_discretisee_implicite}), classical methods for 
optimization problems with constraints can be used. We refer
to~\cite{glowinski-le-lattec-89} for a presentation of the classical Uzawa algorithm, and
to~\cite{bonnans-gilbert-lemarechal-sagastizabal-02} for more advanced
methods. Problem~(\ref{SDE_discretisee_explicite}) can be solved using classical
methods for nonlinear problems, such as the Newton method
(see~\cite{bonnans-gilbert-lemarechal-sagastizabal-02}). We also refer to
Chapter 7 of~\cite{leimkuhler-reich-04} where similar problems are
discussed, for the classical RATTLE and SHAKE schemes used for Hamiltonian dynamics with constraints.

Both schemes are consistent (the discretization error goes to $0$ when the time step $\Delta t$ goes to $0$) with the projected
diffusion~\eqref{eq:Q_proj} (see~\cite{LLV05}). Accordingly,
$\Delta\Lambda_{n+1}$ is a consistent discretization of
$\int_{t_n}^{t_{n+1}} d \Lambda_{t}$
and therefore, it can be proven~\cite{LLV05}:
\[
\lim_{T \to \infty} \lim_{\Delta t \to 0} \frac 1 T
\sum_{n=1}^{T/\Delta t} \Delta\Lambda_{n}=F'(z)
\]
which is the discrete counterpart of the trajectory average~(\ref{eq:F'}). In~\cite{LLV05}, a
variance reduction technique is proposed, which consists in extracting
the bounded variation part $\Delta\Lambda^{\rm f}_n$ of
$\Delta\Lambda_n$ (resorting locally to reversed Brownian
increments). We give some details of an adaptation of this method for evolving constraints in next section.

\subsection{Discretization with evolving constraints}
\label{sec:disc_evolv_const}

When nonequilibrium dynamics are considered, the constraint is stated as $\xi(Q_t)=z(t)$.
The reaction coordinate path is first discretized as $\{ z(0), \dots,
z(t_{N_T}) \}$ where $N_T$ is the number of timesteps. For example, equal time increments can be used, in which
case $\Delta t=\frac{T}{N_T}$ and $t_n =n \Delta t$ (we refer to
Remark~\ref{rem:pract} below for some refinements). The initial conditions $Q_{0}$
are sampled according to $\mu_{\Sigma_0}$. A way to do that is to subsample a long
trajectory of the projected SDE on $\Sigma_0$ (using the schemes~\eqref{SDE_discretisee_implicite} or~\eqref{SDE_discretisee_explicite}). 

The projected SDE on evolving constraints \eqref{e:diff} is then discretized with the scheme \eqref{SDE_discretisee_implicite} or \eqref{SDE_discretisee_explicite}, taking into account the evolution of the constraint:
\begin{equation}
\label{SDE_evolution_implicite}
\left \{ \begin{array}{l}
Q_{n+1} = Q_n - \nabla V(Q_n) \, \Delta t + \sqrt{2 \Delta t \, \beta^{-1}} \, U_n + \Delta\Lambda_{n+1} \, \nabla \xi(Q_{n+1}), \\
\mbox{where $\Delta\Lambda_{n+1}$ is such that $\xi(Q_{n+1}) = z(t_{n+1})$,}
\end{array} \right.
\end{equation}
or
\begin{equation}
\label{SDE_evolution_explicite}
\left \{ \begin{array}{l}
Q_{n+1} = Q_n - \nabla V(Q_n) \, \Delta t + \sqrt{2 \Delta t \, \beta^{-1}}  \, U_n + \Delta\Lambda_{n+1} \, \nabla \xi(Q_n), \\
\mbox{where $\Delta\Lambda_{n+1}$ is such that $\xi(Q_{n+1}) = z(t_{n+1})$.}
\end{array} \right.
\end{equation}

It remains to extract the force part $ \Delta\Lambda^{\rm f}_{n+1} $ from the
discretized Lagrange multiplier $\Delta\Lambda_{n+1}$ (consistently
with~\eqref{e:lambdadec}). We propose two methods. First, this can be
done by simply subtracting the  drift and the martingale part
\begin{equation}
\label{mean_force_evolution}
\Delta\Lambda^{\rm f}_{n+1} = \Delta\Lambda_{n+1} - \frac{z(t_{n+1})-z(t_{n})}{|\nabla \xi(Q_n)|^2} +
\sqrt{  2\Delta t \beta^{-1}} \, \frac{\nabla \xi(Q_{n})}{|\nabla \xi(Q_n)|^2} \cdot U_n.
\end{equation}
Another possibility in the spirit of the variance reduction
techniques used in~\cite{LLV05} can also be used. Consider the following
coupled dynamic with locally time-reversed constraint evolution (written
here for the scheme~\eqref{SDE_evolution_implicite}):
\begin{equation*}
\label{SDEdiscrev}
{Q}^{\rm R}_{n+1} = Q_n - \nabla V(Q_n) \, \Delta t - \sqrt{2\Delta
  t\,\beta^{-1}} \, U_n + \Delta\Lambda^{\rm R}_{n+1} \, \nabla \xi
(Q^{\rm R}_{n+1}),
\end{equation*}
with $\Delta\Lambda^{\rm R}_{n+1}$ such that:
\begin{equation*}
\frac{1}{2}(\xi(Q^{\rm R}_{n+1})+\xi(Q_{n+1}))=\xi(Q_{n}).
\end{equation*}
The position ${Q}^{\rm R}_{n+1} $ is computed as $Q_{n+1}$ in
\eqref{SDE_evolution_implicite}, but with a projection on
$\Sigma_{2\xi(Q_{n}) - \xi(Q_{n+1})}$ instead of $\Sigma_{z(t_{n+1})}$,
and using the Brownian increment $-\sqrt{\Delta t} \, U_{n}$ instead of
$\sqrt{\Delta t} \, U_{n}$. Notice that in case of a constant
increment for the constraints, we have $\xi(Q^{\rm R}_{n+1})=2\xi(Q_{n}) - \xi(Q_{n+1})=z(t_{n-1})$.
The force part $ \Delta\Lambda^{\rm f}_{n+1} $ is then obtained through
\begin{equation}
\label{mean_force_evolution_prime}
\Delta\Lambda^{\rm f}_{n+1} = \frac{1}{2}(\Delta\Lambda_{n+1}
+\Delta\Lambda^{\rm R}_{n+1}  )
\end{equation}
which can be shown to be a consistent time discretization of
$\int_{t_n}^{t_{n+1}}d \Lambda^{\rm f}_{t}$.

\subsection{Computation of free energy using a Feynman-Kac equality}
\label{subsection_discretized_jarz}

The consistent discretization of $Q_t$, and more precisely of
$\int_{t_n}^{t_{n+1}} d\Lambda^{\rm f}_t$, we have obtained in the previous section can now be used
to approximate the work $\W(t)$ defined by~\eqref{eq:work_prac} by
\begin{equation}
\label{eq:num_work}
\left\{
\begin{array}{l}
\W_0=0,\\
\W_{n+1} = \W_{n} + \Frac{z(t_{n+1})-z(t_{n})}{t_{n+1}-t_{n}} \,
\Delta\Lambda^{\rm f}_{n+1},
\end{array}
\right.
\end{equation}
using either the dynamics~(\ref{SDE_evolution_implicite})
or~(\ref{SDE_evolution_explicite}), and the local force part of the Lagrange
multiplier computed by~(\ref{mean_force_evolution}) or~(\ref{mean_force_evolution_prime}).
Averaging over $M$ independent realizations (the corresponding works
being labeled by an upper index $1 \leq m \leq M$), an estimator of the
free energy difference $\Delta F(z(T))$ is, using Theorem~\ref{th:FK},
\begin{equation}
\label{eq:jarz_estimator}
\widehat{\Delta F}(z(T)) = - \beta^{-1} \ln \left ( \frac{1}{M} \sum_{m=1}^M {\rm e}^{-\beta \W_{N_T}^m} \right ).
\end{equation}
The estimator $\widehat{\Delta F}(z(T))$ converges to $\Delta F(z(T))$
as $\Delta t \to 0$ and $M \to +\infty$. It is clear that the estimation
of $\Delta F(z(T))$ by~(\ref{eq:jarz_estimator}) is straightforward to
parallelize since the $(\W_{N_T}^m)_{1 \le m \le M}$ are independent.

Notice that, even in the limit $\Delta t \to 0$, $\widehat{\Delta
  F}(z(T))$ is a biased estimator. Indeed, \linebreak $\exp(-\beta\widehat{\Delta
F}(z(T)))$ is an unbiased estimator of $\exp(-\beta \Delta
F(z(T)))$, and therefore, using the concavity of $\ln$, $\mathbb{E}(\widehat{\Delta F}(z(T)))
\geq \Delta F(z(T))$. Recent works propose corrections to this
systematic bias using asymptotic expansions in the limit $M \to
+\infty$~(see for instance~\cite{RD03,ZW04}). 

\begin{remark}[On practical implementation]\label{rem:pract}
Notice that it may be useful to  adaptively refine the time step over
each stochastic trajectories, using for example the work evolution rate
$(\W_{n}-\W_{n-1})_{n \geq 1}$ as a refinement criterion.

As noticed in~\cite{RD03}, it is also possible to optimize the evolution of the constraint $z(t)$, for
example by minimizing the variance of the results obtained for {\it a
  priori} schedules for the evolving constraint on a small set of preliminary runs.
\end{remark}


\section{Numerical results}
\label{section_res_num}

We present in this section some illustrations of the algorithm we have
described above to compute free energy differences through
nonequilibrium paths. In Section~\ref{sec:toy}, a two-dimensional toy potential $V$ is
used, for which we can compare the results with analytical profiles. A more realistic test case
in Section~\ref{WCA} demonstrates the ability of the method to compute
free energy profiles in presence of a free energy barrier.

Our aim in this section is not to compare the numerical efficiency of
the thermodynamic integration method presented in
Section~\ref{section_TI} (or any other method) with nonequilibrium
computations, since it is difficult to draw {\it general} conclusions about
such comparisons.  However, we compare on a simple example in Section~\ref{sec:toy}, the numerical efficiency of out-of-equilibrium
computations using a few long trajectories or many short trajectories, at a fixed computational
cost.


\subsection{A two-dimensional toy problem}
\label{sec:toy}

\begin{figure}
\begin{center}
\hspace{-2cm}
\includegraphics[width=9cm]{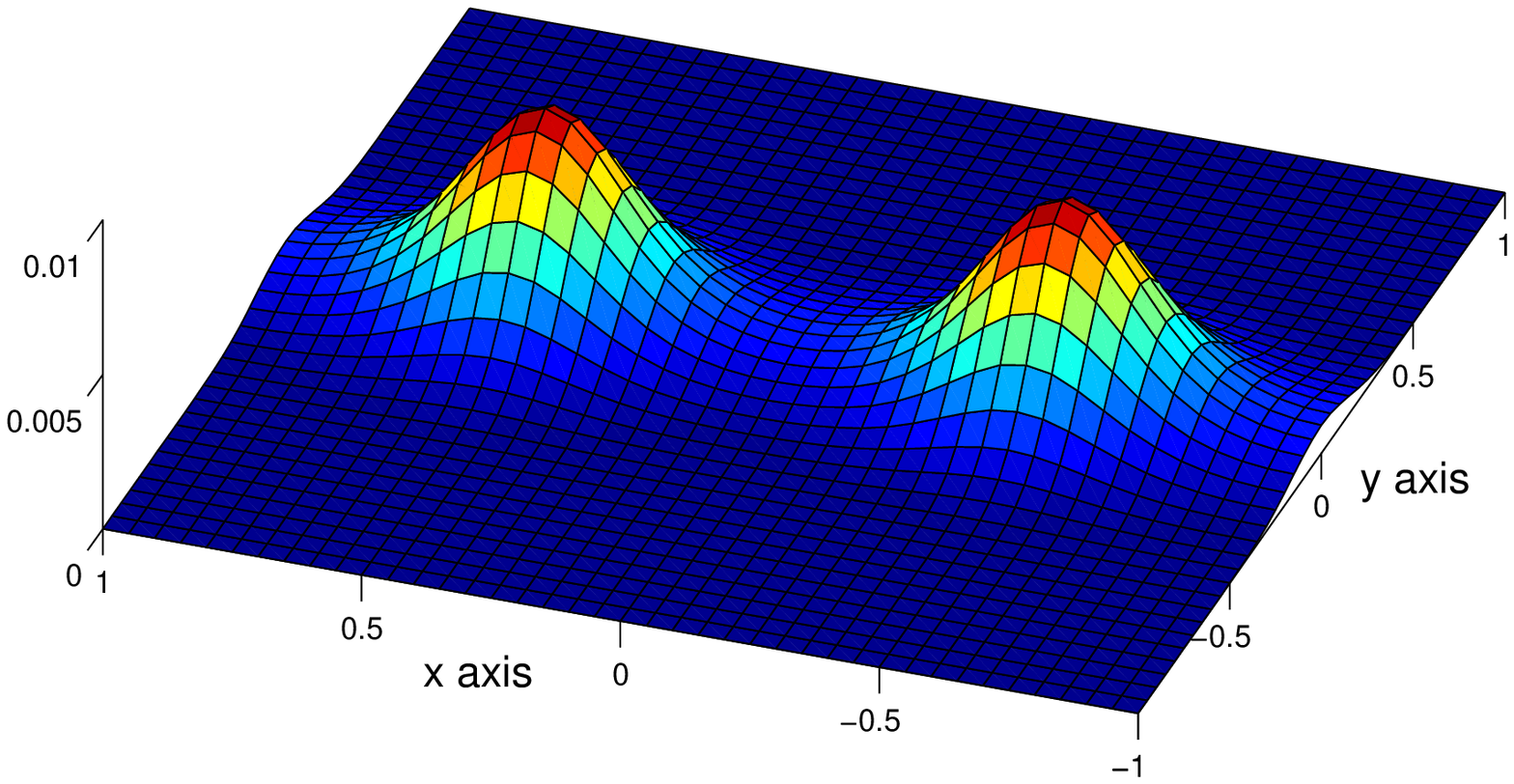}
\hspace{-1cm}
\includegraphics[width=9cm]{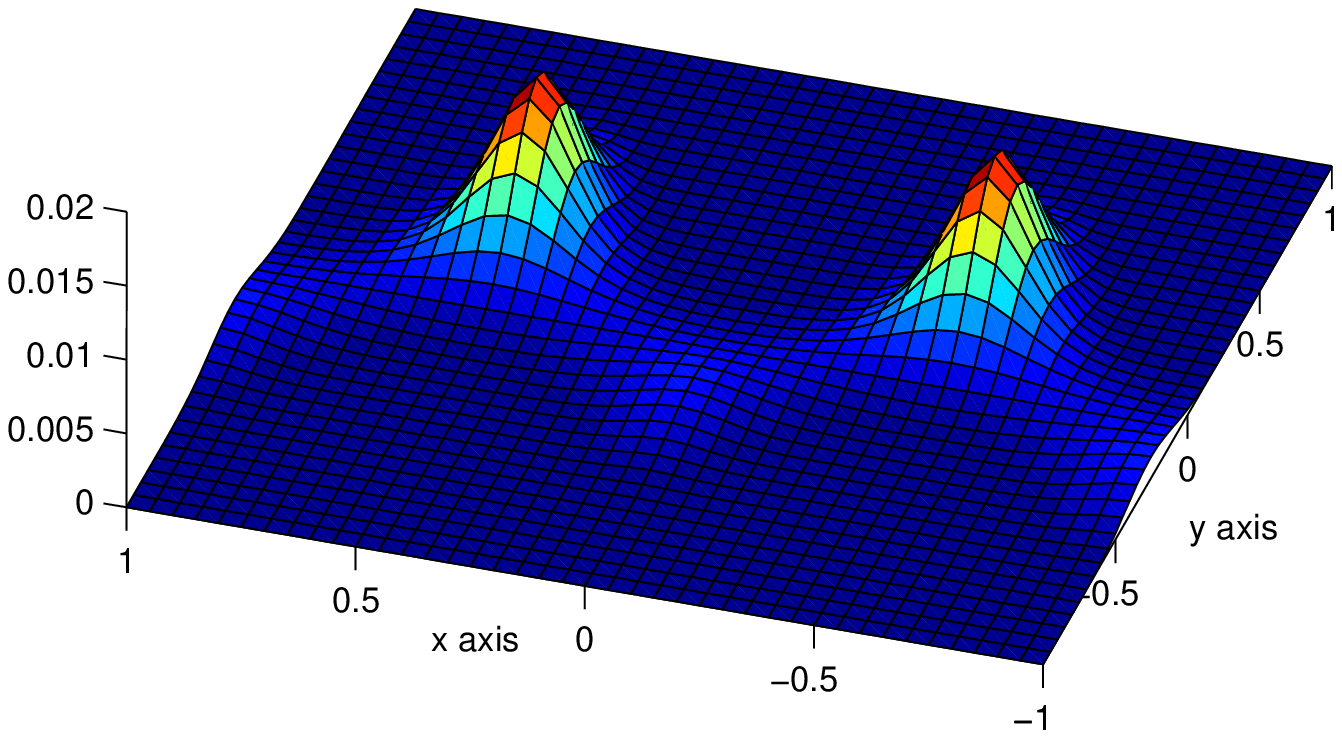}
\end{center}
\caption{Plot of some probability densities corresponding to the
  potential~(\ref{pot_2D}) for $\beta =1$, $d_2=2 \pi^2$, and
  $d_1=0$ on the left or $d_1=10$ on the right. }
\label{densite_2D}
\end{figure}

We consider the two-dimensional potential introduced in~\cite{Voter97}
\begin{equation}
\label{pot_2D}
V(x,y) = \cos(2\pi x) (1+d_1 y) + d_2y^2,
\end{equation}
where $d_1$ and $d_2$ are two positive constants. Some corresponding
Boltzmann-Gibbs probability densities are depicted in Figure~\ref{densite_2D}.

We want to compute the free energy difference profile between the
initial state $x=x_0=-0.5$ and the transition state $x=x_1 = 0$. 
Notice that the saddle point is $(x_1,y_1)=(0,0)$ for $d_1=0$, but
is increasingly shifted toward lower values of $y_1$ as $d_1$ increases. 
We parameterize the transition along the $x$-axis, either with the reaction coordinate
\begin{equation}
\label{coord_2D_naturelle}
\xi(x,y) = \frac{x-x_0}{x_1-x_0},
\end{equation}
or with the reaction coordinate ($n \geq 2$)
\begin{equation}
\label{coord_2D_autre}
\eta_n(x,y) = \frac{1}{2^n-1} \left [ \left ( 1 + \frac{x-x_0}{x_1-x_0}
  \right )^n -1 \right ].
\end{equation}
For these reaction coordinates, the initial state (resp. the transition state) corresponds to a value
of the reaction coordinate $z=0$ (resp. $z=1$). The analytical expression of the free energy difference that we consider
here is, for a reaction coordinate $\nu(x,y)$ (such as $\xi$ or $\eta_n$
defined above)
\begin{equation*}
\Delta F_\nu(z) = - \beta^{-1} \ln \left ( \frac{\int {\rm e}^{-\beta V(x,y)} \delta_{\nu(x,y)-z}  }
{\int {\rm e}^{-\beta V(x,y)} \delta_{\nu(x,y)}  }
\right ),
\end{equation*}
where the distribution $\delta_{\nu(x,y)-z}$ is defined in Remark~\ref{rem:def_F}
above. Notice that even though the initial state $\Sigma_0=\{x=-0.5\}$ and the final
state $\Sigma_1=\{x=0\}$ are the same for the reaction coordinates $\xi$ and
$\eta_n$, the associated free energy differences differ. This is due to the fact that
$\nabla \xi \neq \nabla \eta_n$, and therefore $\delta_{\xi(x,y)-z} \not = \delta_{\eta_n(x,y)-z}$.
More precisely, 
\[
\Delta F_\xi(z) = -\cos (2\pi x_0) + \cos (2\pi x_\xi(z)) + \frac{(d_1)^2}{4d_2} (\cos^2(2\pi x_0) - \cos^2
(2\pi x_\xi(z))),
\]
with \[ x_\xi(z) = x_0 + z(x_1-x_0), \]
and
\begin{align*}
\Delta F_{\eta_n}(z) &= -\cos (2\pi x_0) + \cos (2\pi x_{\eta_n}(z)) +
\frac{(d_1)^2}{4d_2} (\cos^2(2\pi x_0) - \cos^2 (2\pi x_{\eta_n}(z))) \\
&\quad + \frac{n-1}{\beta}\ln \left ( 1 + \frac{x_{\eta_n}(z)-x_0}{x_1-x_0} \right ),
\end{align*}
with
\[ 
x_{\eta_n}(z) = x_0 + ( (2^n-1)z+1)^{1/n}-1)(x_1-x_0).
\]

\begin{figure}
\centering
\hspace{-1cm}
\includegraphics[width=8cm]{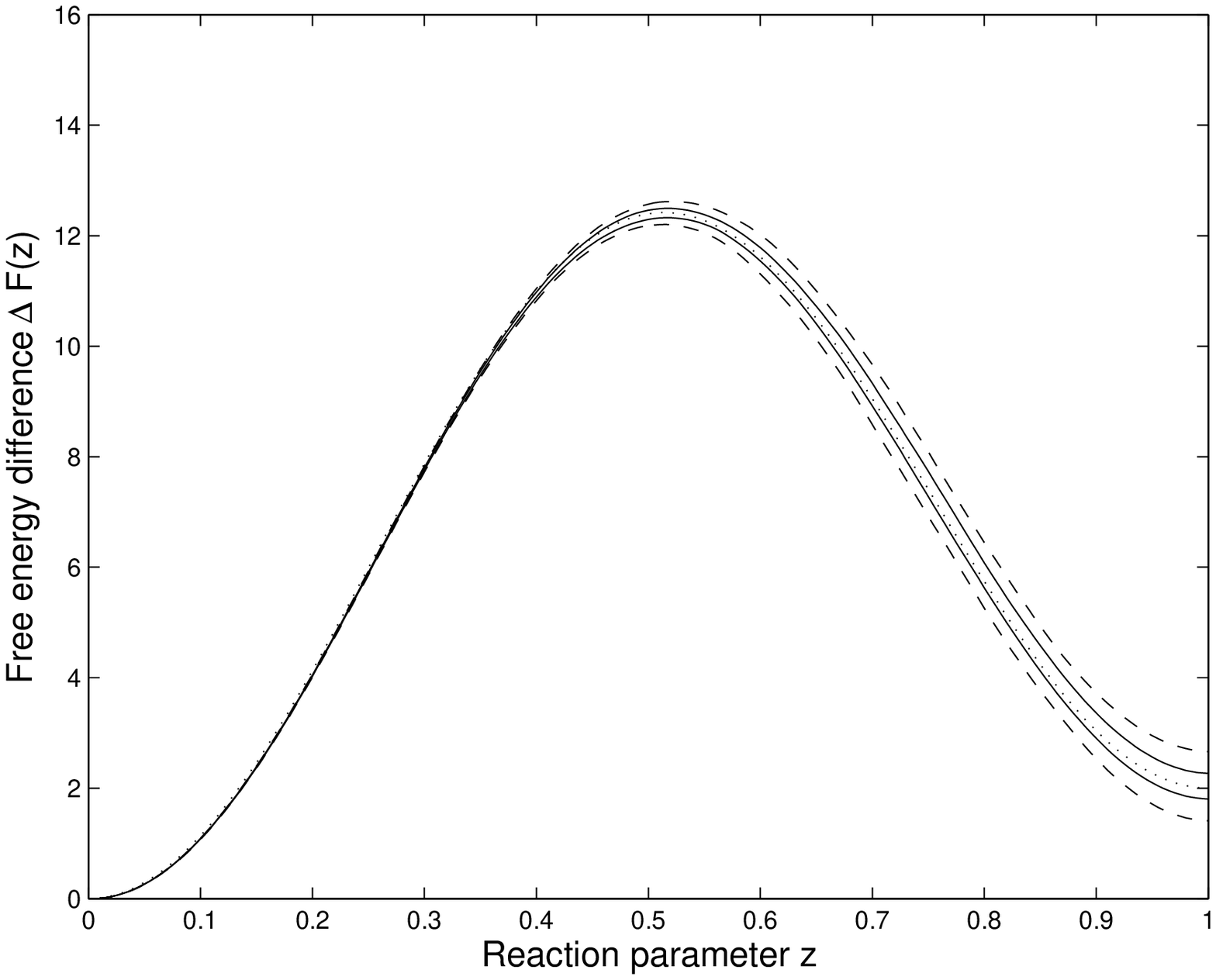}
\includegraphics[width=8cm]{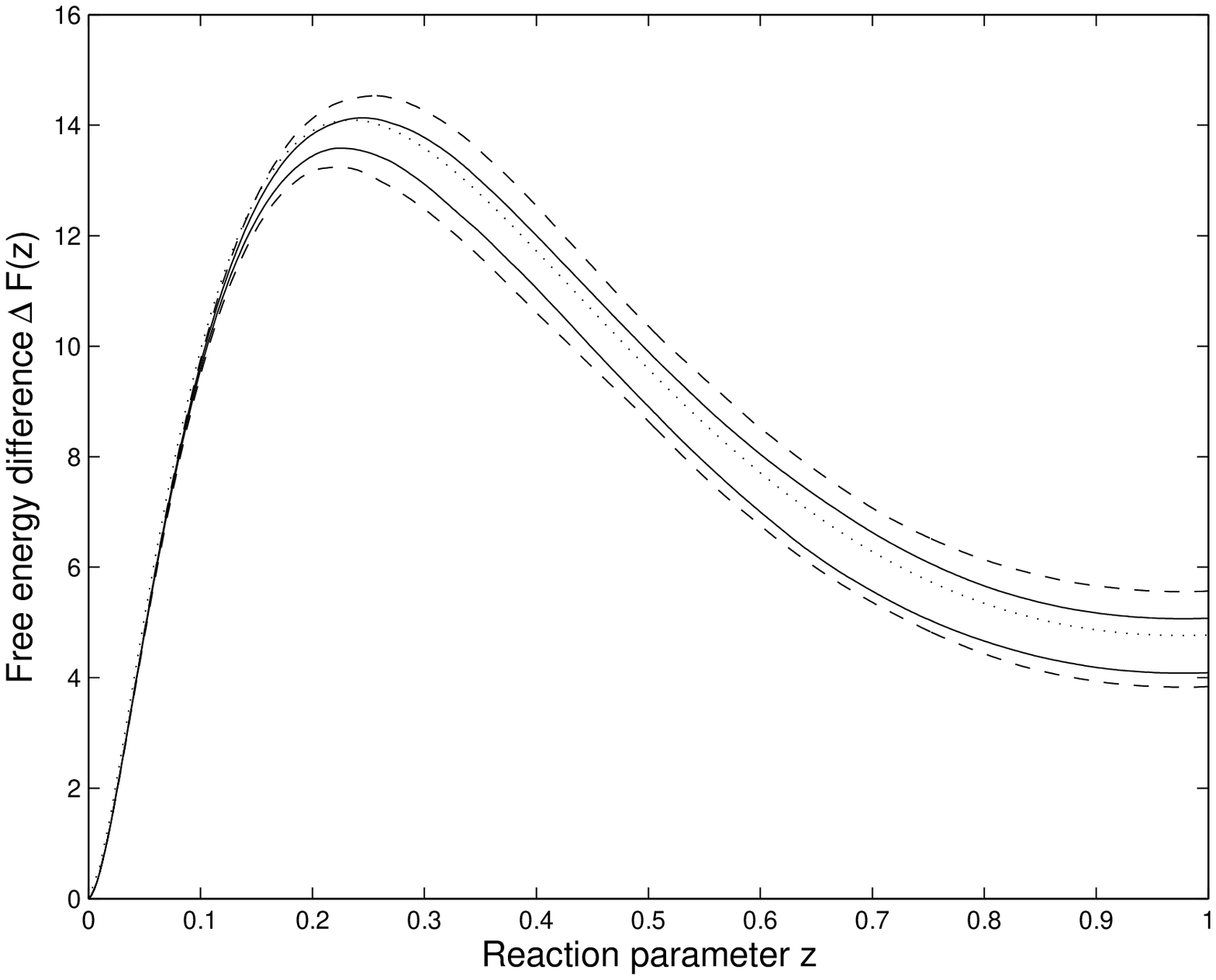}
\caption{Free energy profiles using the potential~(\ref{pot_2D})
  with $\beta=1$, $d_1=30$ and $d_2=2\pi^2$, and the reaction
  coordinate~(\ref{coord_2D_naturelle}) on the left, or the  reaction
  coordinate~(\ref{coord_2D_autre}) with $n=5$ on the right. Analytical
  reference profiles are in dotted
  lines. The dashed lines (resp. the solid lines) represent the upper and lower bound of the
  $95\,\%$ confidence interval (obtained over 100 independent
  realizations) for nonequilibrium computations with $M=10^3$ replicas
  (resp. with $M=10^4$ replicas). The switching time is $T=1$ and the
  time step is $\Delta t = 0.005$ on the left and $\Delta t = 0.0025$ on the right.}
\label{2Dprofiles}
\end{figure}

Free energy profiles for the two reaction coordinates considered here
can then be computed using the discretization proposed in
Section~\ref{subsection_discretized_jarz}. Averaging over several
realizations, error estimates can be proposed: in particular, the
standard deviation can be computed for all intermediate points $z \in
[0,1]$, so that, for all values $z$, a confidence interval around the
empirical mean can be proposed. We represent on Figure~\ref{2Dprofiles} the
analytical profiles, and the
lower and upper bounds of the $95\,\%$ confidence interval for $M=10^3$
and $M=10^4$, using here and henceforth a linear schedule: $z(t)=t/T$. 
The initial conditions are created by subsampling every 100 timesteps a trajectory
constrained to remain on the initial submanifold~$\Sigma_0$. 
As announced above, the profiles obtained with $\eta_n$ and~$\xi$ are not exactly the same, 
though the general shape is preserved. These figures also show that the variance increases 
with~$z$. Therefore, to further test the convergence of the method, it is enough here to 
characterize the convergence of the value for the end point at $z=1$.

We study the convergence of the end value $\Delta F(1)$ computed with
the out-of-equilibrium dynamics with respect to the number of replicas
$M$ and the time step~$\Delta t$, using the
reaction coordinate~(\ref{coord_2D_naturelle}) as an example. 
The results are presented in Table~\ref{results_2D}.  The time step $\Delta t$ does not seem to have any
noticeable influence on the final result, as long as it remains in a
reasonable range. As expected, the error gets smaller as~$M$ 
increases.

In Table~\ref{results_2D}, we also show that, in this particular case, for a fixed
computational cost and provided that the switching time is large
enough\footnote{Of course, this threshold time depends on the system
  under study.},
computing many short trajectories is as efficient as computing a
few longer ones (the mean and the variance are essentially
unchanged). This conclusion also holds for the
more realistic test case presented in next section. The computation of
many trajectories can be straightforwardly and very efficiently parallelized.

We finally mention that we are able to exhibit the bias of the Jarzynski
estimator in this particular case (see Section~\ref{subsection_discretized_jarz} 
and~\cite{ZW04}).
 We observe that the estimator $\widehat{\Delta F}(z(T))$
is generally greater than $\Delta F(z(T))$. More precisely, averaging
over $10^4$ realizations, with the parameters $T=1$ and $\Delta
t=0.005$, we obtain the following 95 $\%$ confidence
intervals for $\widehat{\Delta F}(z(T))$, for various values of $M$: $\widehat{\Delta F}(z(T))=2.0576 \pm 0.0059$ for $M=10^3$, 
$\widehat{\Delta F}(z(T))=2.0095 \pm 0.0026$ for $M=10^4$, and
$\widehat{\Delta F}(z(T))=2.00075\pm 0.0010$ for $M=10^5$.
As expected, the bias goes to zero when $M \to \infty$.

\begin{table}
\centering
\begin{tabular}{cc}
\begin{tabular}{| c c c| c |}
\hline
\hline
 $\Delta t$ & $T$ & $M$ & $\widehat{\Delta F}(z(T))$ \\
\hline
 0.001 & 1 & $10^3$ & 2.056 \ (0.274)\\
 0.0025 & 1 & $10^3$ & 2.033 \ (0.259)\\
 0.005 & 1 & $10^3$  &  2.076 \ (0.286) \\
 0.01 & 1 & $10^3$  &  2.073 \ (0.278) \\
\hline
 0.005 & 1 & $10^3$  &  2.076 \ (0.286) \\
 0.005 & 1 & $10^4$  & 2.014 \ (0.116)  \\
 0.005 & 1 & $10^5$  &  2.001 \ (0.045)  \\
\hline
\hline
\end{tabular}
~~~~~~~~~
\begin{tabular}{| c c c| c |}
\hline
\hline
 $\Delta t$ & $T$ & $M$ & $\widehat{\Delta F}(z(T))$ \\
\hline
 0.005 & 1 & $10^4$  &  2.014 \ (0.116) \\
 0.005 & 10 & $10^3$  & 1.999 (0.029) \\
 0.005 & 100 & $10^2$ & 2.001 (0.025)  \\
 0.005 & 1000 & $10^1$ & 1.997 (0.022)  \\
\hline
\hline
\end{tabular}
\end{tabular}
\caption{\label{results_2D} Free energy differences $\Delta F(1)$
  obtained by nonequilibrium computations for the
  reaction coordinate~(\ref{coord_2D_naturelle}) with $\beta =1$, $d_1=1$  and $d_2=30$. The
  results are presented as follows: $\E\left(\widehat{\Delta F}(z(T)) \right) \quad
  \left(\sqrt{{\rm Var}\left(\widehat{\Delta
          F}(z(T))\right)}\right)$ (the estimates of these quantities
  are obtained by averages over 100 independent runs). The exact value is $\Delta
  F(1)=2$.}
\end{table}

\subsection{Model system for conformational changes influenced by
  solvation}
\label{WCA}

\begin{figure}
\centering
\hspace{-0cm}
\includegraphics[width=7cm]{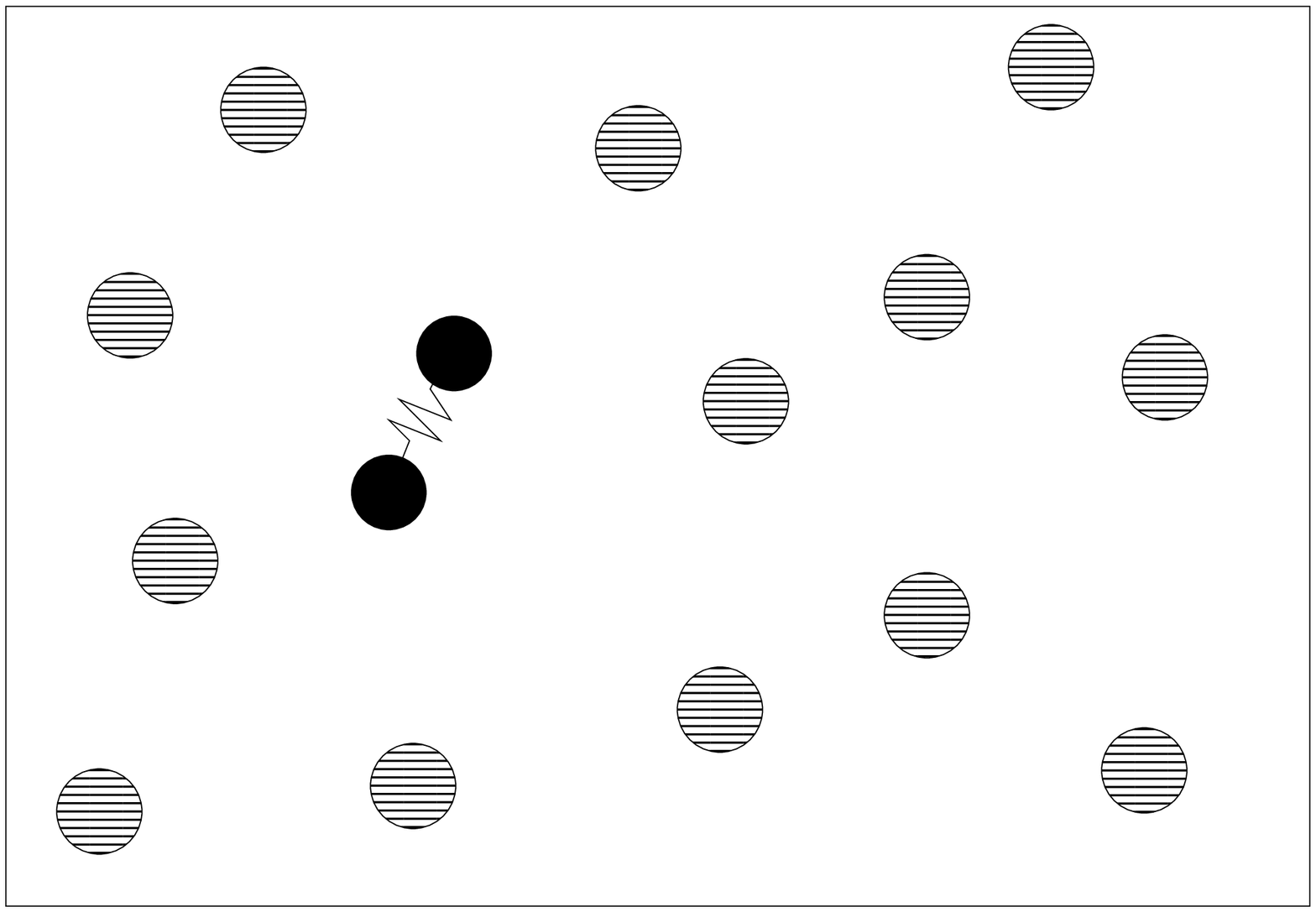}\hfill
\includegraphics[width=7cm]{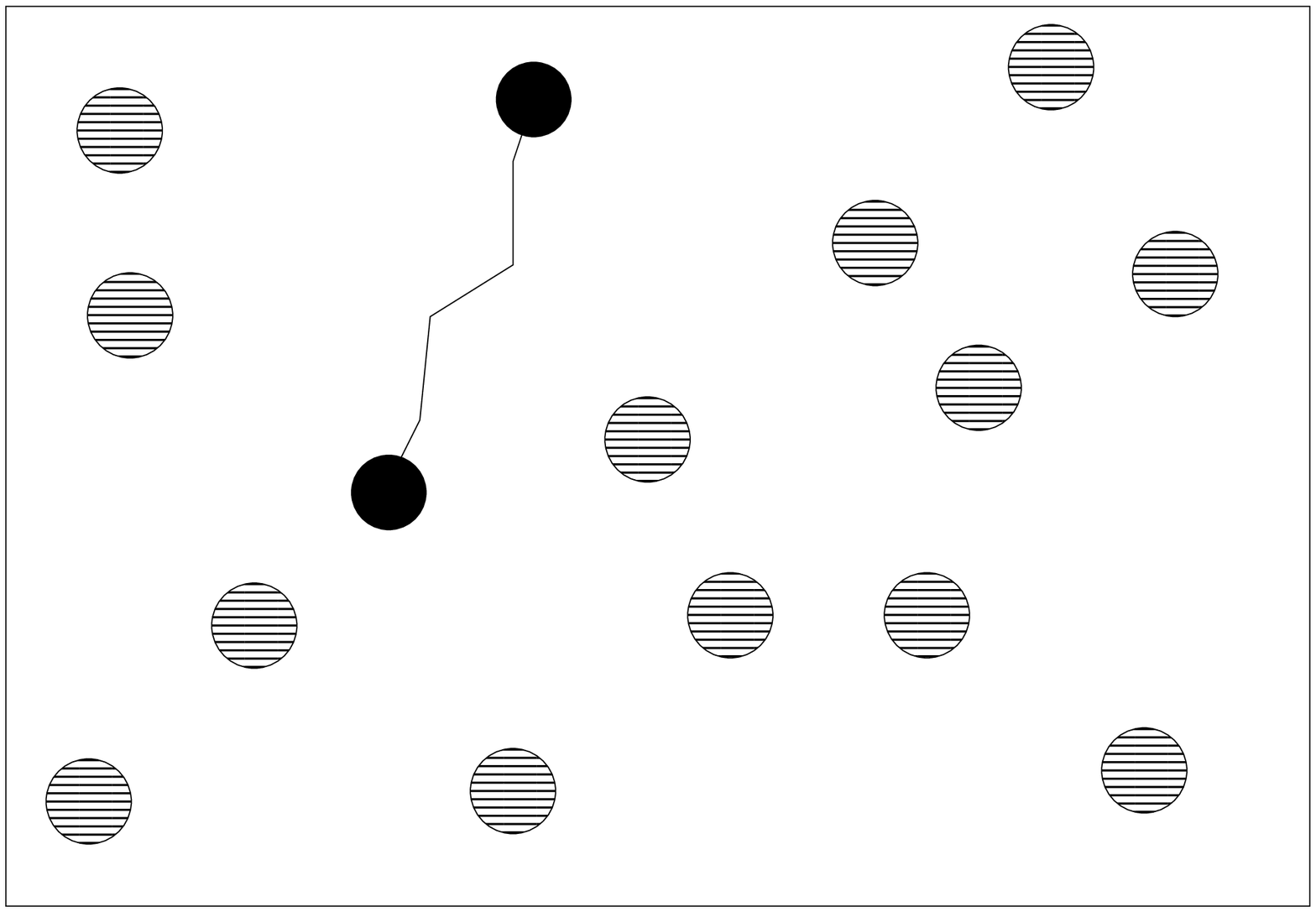}
\caption{\label{WCA_fig} Schematic views of the system, when the dimer is in the compact state (Left), and in the stretched state (Right). The interaction of the particles forming the dimer is described by a double well potential. All the other interactions are of WCA form.}
\end{figure}

We consider a system composed of $N$ particles in a periodic box of side
length $l$, interacting through the purely repulsive WCA pair potential~\cite{DBC99,SBB88}:
\[
V_{\rm WCA}(r) = \left \{ \begin{array}{cl}
4 \epsilon \left [ \left ( \Frac{\sigma}{r} \right )^{12} - \left ( \Frac{\sigma}{r}\right )^6 \right ] + \epsilon & {\rm if \ } r \leq r_0, \\
0 & {\rm if \ } r > r_0,
\end{array} \right.
\]
where $r$ denotes the distance between two particles, $\epsilon$ and $\sigma$ are two positive parameters and $r_0=2^{1/6}\sigma$.
Among these particles, two (numbered 1 and 2 in the following) are
designated to form a dimer while the others are
solvent particles. Instead of the above WCA
potential, the interaction potential between the two particles of the dimer is a double-well potential
\begin{equation}\label{eq:VS}
V_{\rm S}(r) = h \left [ 1 - \frac{(r-r_0-w)^2}{w^2} \right ]^2,
\end{equation}
where $h$ and $w$ are two positive parameters.  
The potential $V_{\rm S}$ exhibits two energy minima, one corresponding
to the compact state where the length of the dimer is $r=r_0$, and one corresponding to
the stretched state where this length is
$r=r_0+2w$. The energy barrier separating both states is $h$. Figure~\ref{WCA_fig} presents a schematic view of the system.

The reaction coordinate used is 
\begin{equation}
\label{coord_reac_diatomic}
\xi(q) = \frac{|q_1-q_2|-r_0}{2w},
\end{equation}
where $q_1$ and $q_2$ are the positions of the particles forming the
dimer. The compact state (resp. the stretched state) corresponds to a
value of the reaction coordinate $z=0$ (resp. $z=1$).

\begin{figure}
\centering
\hspace{-1cm}
\includegraphics[width=8cm]{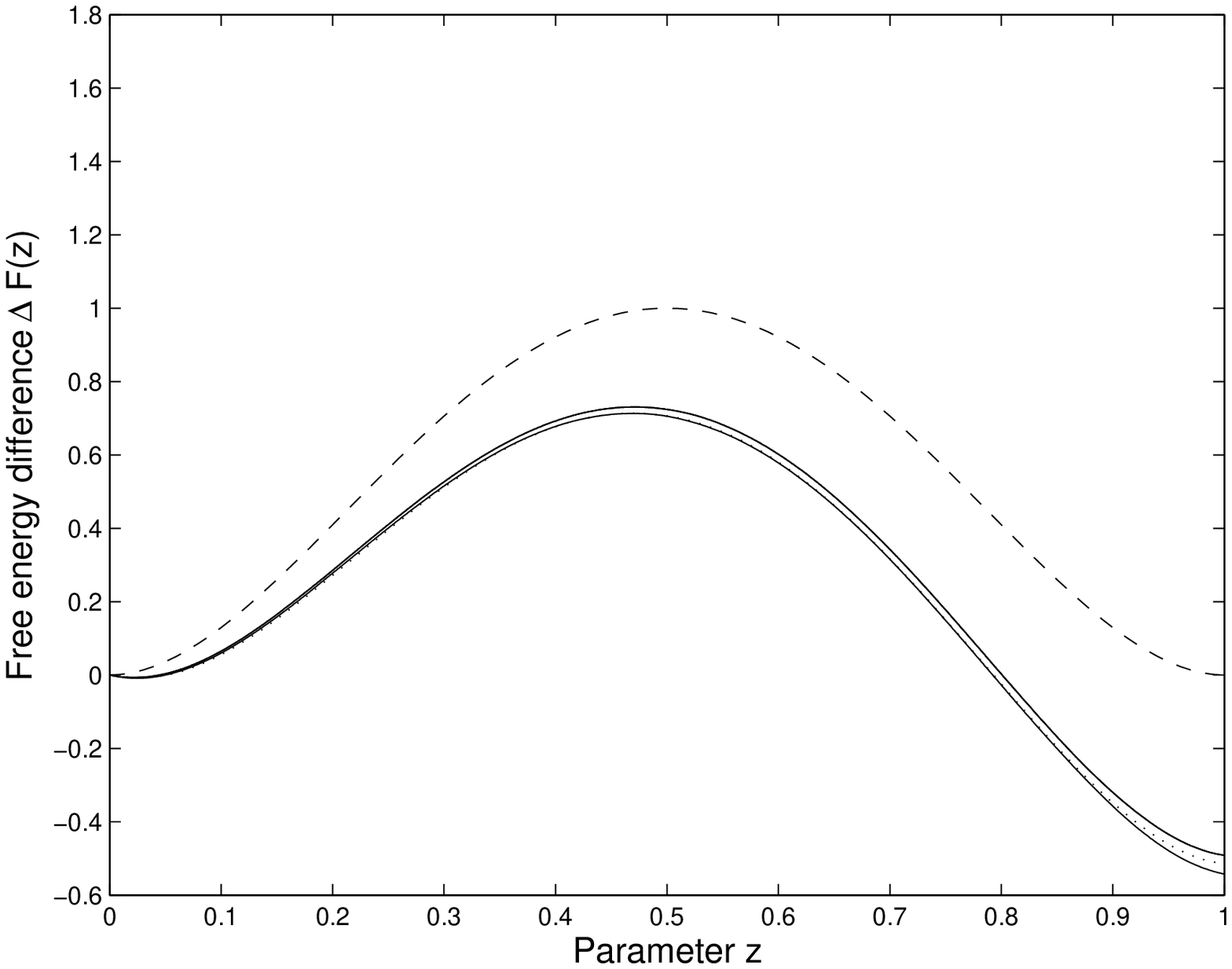}
\includegraphics[width=8cm]{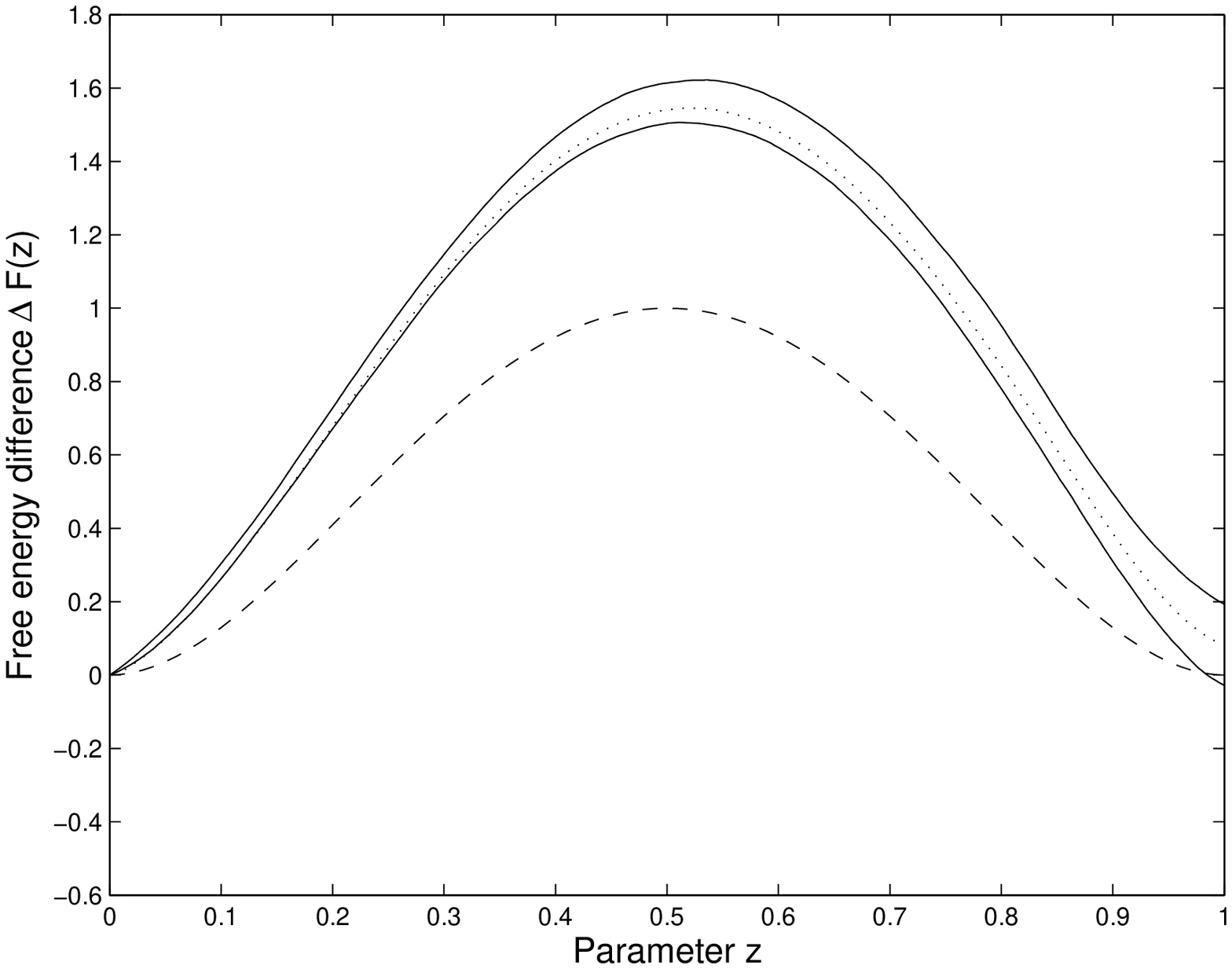}
\caption{Comparison of free energy difference profiles using the reaction
coordinate~(\ref{coord_reac_diatomic}), at low densities ($l=3$) on the
left, and high densities ($l=1.3$) on the right. The
  double well potential $V_S$ is represented in dashed line. The
  reference free energy difference profile  
computed with a very precise thermodynamic integration is represented in
dotted line. We used $N_{\rm TI}=101$ thermodynamic integration points
 (uniformly distributed over $(0,1)$) and averaged the mean force over $M_{\rm TI}=10^7$ configurations for each
fixed value of $z$. The upper and lower bounds of the $95 \, \%$
confidence interval (obtained over 50 independent realizations) for
out-of-equilibrium computations are represented with solid lines. We
used $M=1000$ nonequilibrium trajectories, a switching time
$T=1$, and a timestep~$\Delta t=0.00025$ (left) or~$\Delta t=0.0005$ (right).}
\label{diatomic_ex}
\end{figure}

The parameters used for the simulations are: $\beta=1$, $\epsilon=1$,
$\sigma=1$, $h=1$, $w=0.5$ and $N=16$. We still use a linear schedule: $z(t)=t/T$. The side length $l$ of the simulation box
takes two values: $l=1.3$ (high density state) and $l=3$ (low density state). Figure~\ref{diatomic_ex} presents some plots of the free energy
difference profiles computed using nonequilibrium dynamics, as well as thermodynamic 
integration reference profiles. The results show that nonequilibrium estimates are consistent with thermodynamic 
integration. Our experience on this particular example also shows that it is computationally as
efficient to simulate several short nonequilibrium trajectories
(provided the switching time is not too small, say, $T \sim 1$ in the
units used here, so that the diffusion process can take place), or one
single long trajectory where the switching is done slowly (as already  observed
in Section~\ref{sec:toy}).

The free energy profiles highlight
the relative stabilities of the two conformations of the dimer: at low densities (Figure~\ref{diatomic_ex}, Left) the
stretched conformation has a lower free energy and is thus expected to
be more stable (this can indeed be verified by running long molecular dynamics
trajectories and monitoring the time spent in each conformation). When
the density increases, the compact conformation becomes more and more
likely. At the density considered in~Figure~\ref{diatomic_ex} (Right),
the compact state already has a free energy slightly smaller than the
stretched state. Notice also that the free energy barrier increases as
the density increases, so that spontaneous transitions are less and less
frequent. But since we know here a reaction coordinate, we can enforce
the transition. This prevents us from running and monitoring long
trajectories to get sufficient statistics to compare relative
occurrences of both states. 

\section*{Acknowledgments}
The authors would like to thank the referees for their numerous
comments to improve the paper. The authors would also like to thank the Centre International de Rencontres
Ma\-th\'e\-matiques where this work has been partly performed. TL and GS acknowledge financial support from
Action Concert\'ee Incitative Nouvelles Interfaces des Math\'ematiques,
``Simulation mol\'eculaire'', Minist\`ere de la Recherche,
France. Finally, part
of this work was done while GS was attending the program ``Bridging
Time and Length Scales in Materials Science and Biophysics'' at IPAM
(UCLA).

\appendix
\section{Appendix: The multi-dimensional case}
\label{sec:multi}
In this appendix, we generalize the previous results for nonequilibrium
computation of free energy differences to the case of multi-dimensional
reaction coordinates.

\subsection{Geometric setting and basic notation and formulae.}
We consider a $d$-dimensional system of smooth reaction coordinates $\xi=(\xi_{1},\ldots,\xi_{d}): \R^{3N} \rightarrow \R^d,$
non-singular on an open domain $\mathcal{M} \subset \R^{3N}$
$$\forall q \in \mathcal{M}, \quad {\rm range}(\nabla \xi_{1}(q), \ldots, \nabla \xi_{d}(q) ) =d,$$
and a smooth path of associated coordinates
$$ z=(z_{1},\ldots,z_{d}): [0,T] \rightarrow \R^d.$$
Accordingly, we define for all $t \in [0,T]$ a smooth submanifold of codimension $d$ contained in~$\mathcal{M}$:
$$ \Sigma_{z(t)} = \left\{ q\in \R^{3N},\, \xi(q)=z(t) \right\} \subset \mathcal{M}.$$

In the constraints space $\R^{d}$, coordinates are labeled by Greek
letters and we use the summation convention on repeated indices. In the
configuration space $\R^{3N}$, coordinates are labeled by Latin letters
and we also use the summation convention on repeated indices. We denote by $X
\cdot Y=X_{i}Y_{i}$ the scalar product of two vector fields of
$\R^{3N}$, by $M:N=M_{i,j}N_{i,j}$ the contraction of two
tensor fields of $\R^{3N}$, and by $(X\otimes Y)_{i,j}=X_{i}Y_{j}$ the
tensor product of two vector fields of $\R^{3N}$. 

The $d \times d$ matrix
$$G_{\alpha,\gamma}=\nabla \xi_{\alpha} \cdot \nabla \xi_{\gamma}$$
is the Gram matrix of the constraints. It is symmetric and
strictly positive on $\mathcal{M}$. We denote by
$G^{-1}_{\alpha,\gamma}$ the $(\alpha,\gamma)$ component of $G^{-1}$, the inverse
matrix of $G$. At each point $q \in \mathcal{M}$,
we define the orthogonal projection operator
$$P^{\perp}=G^{-1}_{\alpha,\gamma}\nabla \xi_{\alpha} \otimes \nabla
\xi_{\gamma}$$
onto the normal space to $\Sigma_{\xi(q)}$ and the orthogonal projection operator
$$P=\I-P^{\perp}$$
onto the tangent space to $\Sigma_{\xi(q)}$.
The mean curvature vector field of the submanifold is defined by:
\begin{equation}\label{d:H}   H=-\nabla \cdot \left(  ({\rm det} G)^{1/2} G^{-1}_{\alpha,\gamma}\nabla \xi_{\gamma} \right) ({\rm det} G)^{-1/2}\nabla \xi_{\alpha}
\end{equation}
and satisfies:
$$H_{i}=P_{j,k}\nabla_{j} P_{i,k}.$$

We recall the divergence theorem on submanifolds: for any smooth function
$\phi: \R^{3N}  \to \R^{3N}$ with compact support,
\begin{equation}
\label{eq:surf_div}
\int_{\Sigma_z} {\rm div}_{\Sigma}(\phi)\, d\sigma_{\Sigma_z} = -
\int_{\Sigma_z} H \cdot \phi \,  d\sigma_{\Sigma_z}
\end{equation}
where  ${\rm div}_{\Sigma}(\phi) = P_{i,j} \nabla_{i} \phi_{j}$ denotes
the surface divergence, and $\sigma_{\Sigma_z}$ is the induced Lebesgue
measure on the submanifold $\Sigma_z$ of $\R^{3N}$. 

We will also use the co-area formula: for any smooth function
$\phi: \R^{3N}  \to \R$,
\begin{equation}
\label{eq:co_area}
\int_{\R^{3N}} \phi(q) (\det G(q))^{1/2} dq = \int_{\R^d}
\int_{\Sigma_z} \phi \, d\sigma_{\Sigma_z} \, dz.
\end{equation}

These definitions and formulae are provided with more details in~\cite{LLV05}.

\subsection{Free energy and constrained diffusions for multi-dimensional reaction coordinates}
As in the one-dimensional case, the Boltzmann-Gibbs distribution
restricted on the submanifold $\Sigma_{z}$ is defined by:
$$d \mu_{\Sigma_{z}}=Z_{z}^{-1}\exp(-\beta V )
d\sigma_{\Sigma_{z}},$$
with
$$ Z_{z}=\int_{\Sigma_{z}} \exp(-\beta V ) d\sigma_{\Sigma_{z}}.$$
The associated free energy is:
$$ F(z)=-\beta^{-1}\ln \left(Z_{z} \right).$$
\begin{remark}[On the definition of the free energy: the
  multi-dimensional case]
\label{rem:def_F_multi}
As in the one-dimensional case (see Remark~\ref{rem:def_F}), if the particles initially evolve in a potential~$V$, the classical
definition of the free energy is as above, but with $V$ replaced by an
effective potential
$V+\beta^{-1} \ln \left(({\rm det} G)^{1/2}\right)$. The computation of
the gradient of this potential in the dynamics then involves second-order derivatives of $\xi$, which can be approximated in practice by finite differences (see~\cite{LLV05}).
\end{remark}

For any $1 \le \alpha \le d$, we now introduce the local mean force along $\nabla \xi_\alpha$ (which generalizes~(\ref{eq:loc_force})):
\begin{equation}\label{eq:f_alpha} 
f_{\alpha}= G^{-1}_{\alpha,\gamma} \nabla \xi_\gamma \cdot \left(\nabla
  V+\beta^{-1} H\right).
\end{equation}
As in the one-dimensional case (see Equation~(\ref{eq:mean_force_mu})), we obtain the derivative of the mean
force by averaging the local mean force:
\begin{Pro}\label{prop:F_multi}
The derivative of the free energy $F$ with respect to $z_\alpha$ is given by:
\begin{equation*}
\nabla_{\alpha}F(z)=\int_{\Sigma_{z}}\!\!\!\!f_{\alpha} \, \, d\mu_{\Sigma_z}.
\end{equation*}
\end{Pro}
Proposition~\ref{prop:F_multi} is a corollary of
\begin{Lem}\label{l:evo} For any test function $\ph$ with compact support in $\mathcal{M}$, we have:
\begin{equation*}
\nabla_{\alpha} \left( \int_{\Sigma_{z}} \!\!\!\! \ph \exp( -\beta V )
d\sigma_{\Sigma_{z}}\right) = \int_{\Sigma_{z}} \!\!\!\! \left(
  G^{-1}_{\alpha,\gamma} \nabla \xi_{\gamma} \cdot \nabla \ph-\beta
  f_{\alpha} \ph\right) \exp( -\beta V ) d\sigma_{\Sigma_{z}}.
\end{equation*}
\end{Lem}
\begin{proof} It is enough to prove the formula in the case $V=0$, up to
  a modification of the test function $\varphi$. For any test function $g : \R
  \to \R$ with compact support, we have (using successively an integration by parts on
  $\R$, the co-area formula~(\ref{eq:co_area}), an integration by parts on
  $\R^{3N}$, and finally again~(\ref{eq:co_area})):
\begin{align*}
\int_{\R^{d}} g(z_{\alpha}) \nabla_{\alpha}\left(
     \int_{\Sigma_{z}} \!\!\!\! \ph d\sigma_{\Sigma_{z}} \right) \, dz &= - \int_{\R^{d}}  \int_{\Sigma_{z}} \!\!\!\! g'(z_{\alpha}) \ph \, d\sigma_{\Sigma_{z}}dz,\\
&= -\int_{\R^{3N}} \!\! g' \circ \xi_{\alpha} \, \ph \,  \pare{{\rm det}
  G}^{1/2}\, dq,\\
&= -\int_{\R^{3N}} \!\! G^{-1}_{\alpha,\gamma} \nabla \xi_\gamma \cdot \nabla (g \circ \xi_{\alpha}) \,
 \, \ph \,  \pare{{\rm det} G}^{1/2}\, dq,\\
&= \int_{\R^{3N}} \!\! g \circ \xi_{\alpha} \nabla \cdot\left(
  G^{-1}_{\alpha,\gamma} \, \nabla \xi_{\gamma} \, \ph \, \pare{{\rm det} G}^{1/2}\right)\, dq,\\
& =  \int_{\R^{d}} g(z_{\alpha}) \int_{\Sigma_{z}} \!\!\!\! \nabla
\cdot \left( G^{-1}_{\alpha,\gamma}\nabla \xi_{\gamma}\, \ph  \pare{{\rm
      det} G}^{1/2}\right)\pare{{\rm det}
  G}^{-1/2}\,d\sigma_{\Sigma_{z}} \, dz,
\end{align*}
which gives the result using the expression~\eqref{d:H} of the mean
curvature vector $H$.
\end{proof}

We now define the constrained diffusion (which generalizes (\ref{e:diff})):
\begin{equation}\label{e:diffmulti}
\begin{system}
Q_{0}  &\sim & \mu_{\Sigma_{z(0)}},\\
d Q_{t} & = & -P(Q_{t})\nabla V (Q_{t}) dt+\sqrt{2\beta^{-1}} P(Q_{t})
\circ d B_{t}+\nabla \xi_{\alpha}(Q_{t})d \Lambda^{\rm ext}_{\alpha,t}, \\
d \Lambda^{\rm ext}_{\alpha,t}&=&G^{-1}_{\alpha,\gamma}(Q_{t})
z'_{\gamma}(t)dt, \qquad \forall 1 \le \alpha \le d.
\end{system}
\end{equation}
The stochastic process $Q_{t}$ can be characterized by the following property:
\begin{Pro} The process $Q_t$ solution to~\eqref{e:diffmulti} is the
  only It\^o process satisfying for some adapted It\^o processes
  $(\Lambda_{1,t},\ldots,\Lambda_{d,t})_{t\in[0,T]}$ with values in $\R^d$:
\begin{equation*}
\begin{system}
Q_{0}  &\sim & \mu_{\Sigma_{z(0)}},\\
d Q_{t} & = & -\nabla V (Q_{t}) dt+\sqrt{2\beta^{-1}} d B_{t}+\nabla \xi_{\alpha}(Q_{t})d \Lambda_{\alpha,t}, \\
\xi(Q_{t})&=& z(t).
\end{system}
\end{equation*}
Moreover, the process $(\Lambda_{\alpha,t})_{t\in[0,T]}$ can be
decomposed as
 $$ \Lambda_{\alpha,t}=\Lambda^{\rm m}_{\alpha,t}
+\Lambda^{\rm f}_{\alpha,t}+\Lambda^{\rm ext}_{\alpha,t},$$
with the martingale part
$$d \Lambda^{\rm m}_{\alpha,t}= -\sqrt{2\beta^{-1}}
G^{-1}_{\alpha,\gamma} \nabla \xi_\gamma(Q_t) \cdot dB_{t},$$
the local force part (see~(\ref{eq:f_alpha}) for the definition of $f_\alpha$)
\begin{equation*}
d \Lambda^{\rm f}_{\alpha,t}=f_{\alpha}(Q_{t}) dt,
\end{equation*}
and the external forcing (or switching) term
$$ d \Lambda^{\rm ext}_{\alpha,t}= G^{-1}_{\alpha,\gamma}(Q_{t})
z'_{\gamma}(t)dt.$$
\end{Pro}
The proof consists in computing $d\xi(Q_t)$ by It\^o's calculus and
identifying the bounded variation and the martingale parts of the
stochastic processes.

\subsection{The Feynman-Kac fluctuation equality}

Theorem~\ref{th:FK} is generalized as:
\begin{The}[Feynman-Kac fluctuation equality]\label{th:FK_multi}
Let us define the nonequilibrium work exerted on the diffusion $Q_t$
solution to~\eqref{e:diffmulti} by:
\begin{equation*}\W(t)=\int_{0}^{t}f_{\alpha}(Q_{s})z'_{\alpha}(s)\,ds=\int_{0}^{t}z'_{\alpha}(s)d
  \Lambda^{\rm f}_{\alpha,s}.
\end{equation*}
Then, we have the following fluctuation equality: for any test function
$\ph$, and $\forall t \in [0,T]$,
\begin{equation}\label{eq:FK_ph_multi}
\frac{Z_{z(t)}}{Z_{z(0)}}\int_{\Sigma_{z(t)}}\ph \, d\mu_{\Sigma_{z(t)}}=\E\pare{\ph(Q_{t}){\rm e}^{-\beta \W(t)}   }.
\end{equation}
In particular, we have the work fluctuation identity: $\forall t \in [0,T]$,
\begin{equation}\label{eq:FK_multi}
\Delta F(z(t))=F(z(t))-F(z(0))=-\beta^{-1}\ln\pare{ \E\pare{{\rm
      e}^{-\beta \W(t)}   }    }.
\end{equation}
\end{The}
\begin{proof}
For any $s \in [0,T]$ and $x \in  \mathcal{M}$, let us introduce
$(Q^{s,x}_t)_{t \in [s,T]}$, the stochastic process satisfying the
SDE~\eqref{e:diffmulti}, starting from $x$ at time $s$:
\begin{equation}\label{e:diffmulti_cond}
\begin{system}
Q^{s,x}_{s}  &= & x,\\
d Q^{s,x}_{t} & = & -P(Q^{s,x}_{t})\nabla V (Q^{s,x}_{t}) dt+\sqrt{2\beta^{-1}} P(Q^{s,x}_{t})
\circ d B_{t}+\nabla \xi_{\alpha}(Q^{s,x}_{t})d \Lambda^{\rm ext}_{\alpha,t}, \\
d \Lambda^{\rm ext}_{\alpha,t}&=&G^{-1}_{\alpha,\gamma}(Q^{s,x}_{t})
z'_{\gamma}(t)dt, \qquad \forall 1 \le \alpha \le d.
\end{system}
\end{equation}
Notice that for any $s \in [0,T]$, there is an open
neighborhood $(s^{-},s^{+})\times \mathcal{M}_{s}$ of
$(s,\Sigma_{z(s)})$ in $\mathbb{R} \times \mathcal{M}$ such that the
diffusion $(Q^{s,x}_t)_{t \in [s,T]}$ remains in~$\mathcal{M}$ almost surely. This holds since this
process satisfies $d\xi(Q^{s,x}_t)=z'(t)\, dt$ and therefore $\xi(Q^{s,x}_{t})=\xi(x)+z(t)-z(s)$. This gives usual
regularity assumptions sufficient to get a backward semi-group ($t$
being from now on fixed in~$(0,T)$ and $s$ varying in $[0,t]$):
$$ u(s,x)=\mathbb{E} \left( \ph
(Q^{s,x}_{t}) \exp\left(-\beta \int_{s}^{t}f_{\alpha}(Q^{s,x}_{r})z'_{\alpha}(r)\,dr\right)  \right),$$
satisfying the following partial differential equation (PDE) on $(s^{-},s^{+})\times \mathcal{M}_{s}$:
$$ \partial_{s} u = -L_{s}(u(s,.))+\beta z'_{\alpha}(s) f_{\alpha} u, $$
where $L_{s}$ is the generator of the diffusion $Q_t$ solution to~\eqref{e:diffmulti}:
$$ L_{s}=\beta^{-1} P : \nabla ^{2} -P \nabla V \cdot \nabla +\beta^{-1} H \cdot\nabla+z'_{\gamma}(s)G^{-1}_{\alpha,\gamma}  \nabla \xi_{\alpha}\cdot\nabla.$$
Now, using Lemma~\ref{l:evo}, we have:
\begin{eqnarray*}\lefteqn{
\frac{d}{ds} \int_{\Sigma_{z(s)}} \!\!\!\! u(s,.) \exp( -\beta V ) d\sigma_{\Sigma_{z(s)}}}&&\\
&=& \int_{\Sigma_{z(s)}} \!\!\!\! \left(-L_{s}(u(s,.))+ z'_{\alpha}(s)G^{-1}_{\alpha,\gamma} \nabla \xi_{\gamma} \cdot\nabla u(s,.)\right)\exp( -\beta V ) d\sigma_{\Sigma_{z(s)}},\\
&=&-\int_{\Sigma_{z(s)}} \!\!\!\! \left( \beta^{-1} P : \nabla ^{2}u(s,.) -P \nabla V\cdot\nabla u(s,.)+\beta^{-1} H \cdot\nabla u(s,.) \right)\exp( -\beta V ) d\sigma_{\Sigma_{z(s)}},\\
&=&-\beta^{-1} \int_{\Sigma_{z(s)}} \!\!\!\! \Big( {\rm div}_{\Sigma}\left(\nabla u(s,.) \exp( -\beta V )\right)+  H \cdot\nabla u(s,.)\exp( -\beta V ) \Big) d\sigma_{\Sigma_{z(s)}},\\
&=&0,
\end{eqnarray*}
by the divergence theorem~(\ref{eq:surf_div}). Therefore $$\int_{\Sigma_{z(t)}} \!\!\!\!
u(t,.) \exp( -\beta V ) d\sigma_{\Sigma_{z(t)}}=\int_{\Sigma_{z(0)}}
\!\!\!\! u(0,.) \exp( -\beta V ) d\sigma_{\Sigma_{z(0)}},$$ 
which yields
$$\int_{\Sigma_{z(t)}} \!\!\!\!
\ph \exp( -\beta V ) d\sigma_{\Sigma_{z(t)}}=Z_{z(0)}
\mathbb{E} \left( \ph
(Q_{t}) \exp\left(-\beta
  \int_{0}^{t}f_{\alpha}(Q_{r})z'_{\alpha}(r)\,dr\right) \right),$$
where $Q_t$ satisfies~\eqref{e:diffmulti}.
This proves~(\ref{eq:FK_ph_multi}), and~(\ref{eq:FK_multi}) is obtained by taking $\ph=1$.
\end{proof}

\subsection{The numerical scheme}

The adaptation of the algorithm we propose for the one-dimensional case
to the multi-dimensional case is straightforward. Indeed, the
generalizations of schemes~(\ref{SDE_evolution_implicite})
and~(\ref{SDE_evolution_explicite}) to the multi-dimensional case are,
respectively:
\begin{equation*}
\label{SDE_evolution_implicite_multi}
\left \{ \begin{array}{l}
Q_{n+1} = Q_n - \nabla V(Q_n) \, \Delta t + \sqrt{2 \Delta t \, \beta^{-1}} \, U_n + \Delta\Lambda_{\alpha,n+1} \, \nabla \xi_\alpha(Q_{n+1}), \\
\mbox{where $(\Delta\Lambda_{\alpha,n+1})_{1 \le \alpha \le d}$ is such that $\xi(Q_{n+1}) = z(t_{n+1})$,}
\end{array} \right.
\end{equation*}
\begin{equation*}
\label{SDE_evolution_explicite_multi}
\left \{ \begin{array}{l}
Q_{n+1} = Q_n - \nabla V(Q_n) \, \Delta t + \sqrt{2 \Delta t \, \beta^{-1}}  \, U_n + \Delta\Lambda_{\alpha,n+1} \, \nabla \xi_\alpha(Q_n), \\
\mbox{where $(\Delta\Lambda_{\alpha,n+1})_{1 \le \alpha \le d}$ is such that $\xi(Q_{n+1}) = z(t_{n+1})$.}
\end{array} \right.
\end{equation*}
The force part $\Delta \Lambda^{\rm f}_{\alpha,n}$ of $\Delta
\Lambda_{\alpha,n}$ is obtained by similar procedures as those described
in Section~\ref{sec:disc_evolv_const}. For example, the generalization
of~(\ref{mean_force_evolution}) is:
\begin{equation*}
\label{mean_force_evolution_multi}
\Delta\Lambda^{\rm f}_{\alpha,n+1} = \Delta\Lambda_{\alpha,n+1} -
G^{-1}_{\alpha,\gamma}(Q_n) \left(z_\gamma(t_{n+1})-z_\gamma(t_{n})\right) +
\sqrt{  2\Delta t \beta^{-1}} G^{-1}_{\alpha,\gamma} \nabla \xi_\gamma(Q_n) \cdot U_n.
\end{equation*}
The generalization of~(\ref{mean_force_evolution_prime}) is also straightforward.

Now, the estimator $\widehat{\Delta F}(z(T))$ of the free energy difference
$\Delta F(z(T))$ is given by~(\ref{eq:jarz_estimator}), with the
following approximation of the work $\W(t)$:
\begin{equation*}
\label{eq:num_work_multi}
\left\{
\begin{array}{l}
\W_0=0,\\
\W_{n+1} = \W_{n} + \Frac{z_\alpha(t_{n+1})-z_\alpha(t_{n})}{t_{n+1}-t_{n}} \,
\Delta\Lambda^{\rm f}_{\alpha,n+1},
\end{array}
\right.
\end{equation*}
which generalizes~(\ref{eq:num_work}). Notice that
Remark~\ref{rem:pract} also holds
for a multi-dimensional reaction coordinate.

\end{document}